\documentclass[journal]{IEEEtran}

\usepackage[utf8]{inputenc}
\usepackage[T1]{fontenc}

\usepackage{cite}
\usepackage{amsmath,amssymb,amsfonts}
\usepackage{algorithmic}
\usepackage{graphicx}
\usepackage{textcomp}
\usepackage{xcolor}
\usepackage{amsthm}
\usepackage{tikz} 
\usetikzlibrary{positioning} 
\usetikzlibrary{calc,shapes} 
\usepackage{tikz-3dplot}
\newtheorem{theorem}{Theorem}[section]
\newtheorem{lemma}[theorem]{Lemma}
\newtheorem{proposition}[theorem]{Proposition}
\newtheorem{remark}[theorem]{Remark}

\ifCLASSINFOpdf
\else
\fi
\begin{document}
%
\title{Exploiting Multiple Polarizations in Extra Large Holographic MIMO}
%
%
%

\author{Adrian~Agustin,~\IEEEmembership{Senior~Member,~IEEE,}
        and~Xavier~Mestre,~\IEEEmembership{Senior~Member,~IEEE}

\thanks{This work is supported by the grant from the Spanish ministry of economic affairs and digital transformation and of the European Union – NextGenerationEU UNICO-5G I+D/AROMA3D-Earth (TSI-063000-2021-69), by Grant 2021 SGR 00772 funded by the Universities and Research Depart. from Generalitat Catalunya, and by Spanish Government through
project 6G AI-native Air Interface (6G-AINA, PID2021-128373OB-I00
funded by MCIN/AEI/ 10.13039/501100011033) and by ”ERDF A way of
making Europe} Centre Tecnològic de Telecomunicacions de Catalunya (CTTC), Castelldefels.}

%
%

\markboth{Submitted to IEEE Transactions on Wireless Communications}{Agustin \MakeLowercase{\textit{et al.}}}
%



\maketitle

\begin{abstract}
The proliferation of large multi-antenna configurations operating in high frequency bands has recently challenged the conventional far-field, rich-scattering paradigm of wireless channels. Extra large antenna arrays must usually work in the near field and in the absence of multipath, which are far from traditional assumptions in conventional wireless communication systems. The present study proposes to analyze the spatial multiplexing capabilities of large multi-antenna configurations under line-of-sight, near field conditions by considering the use of multiple orthogonal diversities at both transmitter and receiver. The analysis is carried out using a holographic approximation to the problem, whereby the number of radiating elements is assumed to become large while their separation becomes asymptotically negligible. This emulates the operation of a continuous aperture of infinitesimal radiating elements, also recently known as \textit{holographic} surfaces. The present study characterizes the asymptotic MIMO channel as seen by extra large uniform linear and planar arrays, as well as their associated achievable rates assuming access to perfect channel state information (CSI). It is shown, in particular, that for a given distance between the receiver and the center of the array and a given signal quality, there exists an optimum dimension of the multi-antenna surface that maximizes the spectral efficiency.
\end{abstract}


%
\IEEEpeerreviewmaketitle

\section{Introduction}
%
%
%
%

\IEEEPARstart{R}{}ecent trends in multi-antenna technology have prompted an important increase in the number of radiating elements, allowing an impressive enhancement of the advantages of antenna arrays and multiple-input multiple-output (MIMO) technologies. This evolution has motivated the development of massive MIMO (mMIMO) architectures \cite{marzetta2016,bjornson17}, leading to an unprecedented enhancement of spatial multiplexing capabilities. The paradigm has recently evolved towards the eXtra Large MIMO (XL-MIMO) concept, also referred to as Extra Large Antenna Arrays (ELAA), which scale up the system dimensions even further, up to the point where the communication link distance is comparable in magnitude to the spatial dimensions of the array. This paradigm poses new challenges in a diverse number of areas, ranging from hardware design, antenna configuration (whether continuous or discrete aperture) or signal processing algorithms. 

In particular, the enlargement of the array aperture has a profound impact on the assumptions commonly made in traditional wireless communication systems, which are usually designed to operate in the \textit{far-field} region. In XL-MIMO, new electromagnetic (EM) characteristics must be considered, as communication can occur in the \textit{near-field} region, where the spherical propagation wave can no longer be approximated by a planar wave. Consequently, the channel models applicable to XL-MIMO differ from those traditionally used for conventional mMIMO. The \textit{near-field} paradigm presents novel technical opportunities, such as the application of spatial filtering to focus energy on a compact region (i.e., \textit{beamfocusing}) rather than merely directing it along a specific direction \cite{Bjornson19, ramezani2023}. This capability significantly enhances system throughput in wireless networks by drastically reducing interference between terminals, a feature that is expected to play a pivotal role in shaping the future 6G standard.

In parallel with this, wireless systems are moving to higher frequencies in order to exploit the abundant available spectrum, which allow for higher bandwidths and faster data rates \cite{Dahlman24, lin2023}. Higher frequencies also facilitate the deployment of more compact and smaller radiation elements, a fact that has motivated the proliferation of studies on \emph{holographic} surfaces \cite{Dardari21,Huang20}, also known as continuous aperture architectures \cite{Sayeed10} or large intelligent surfaces \cite{Hu18}. These configurations involve an extremely large number of antennas placed with infinitesimally small distances among them. In this framework, the array aperture is viewed as a continuous surface of radiating elements, capable of manipulating electromagnetic waves at the most fundamental physical level, effectively creating "holograms" of electromagnetic fields \cite{Pizzo22}.

Another important consequence of moving towards higher transmission frequencies are the higher propagation losses that are experimented by electromagnetic waves. These losses ultimately contribute to the reduction of multipath (due to the more limited capability of signals to reflect, scatter or diffract off obstacles), so that the Line-of-Sight (LoS) component becomes the dominant or even the only communication path. This is a significant change of paradigm with respect to more conventional MIMO configurations, and many questions still remain unanswered regarding the performance limits of MIMO systems in the absence of a rich scattering environment. 

\subsection{Degrees of freedom for spatial multiplexing (spatial modes)}
Spatial multiplexing refers to the possibility to support the simultaneous transmission of multiple symbol streams in a MIMO wireless system, also referred to as \textit{spatial modes} or spatial degrees of freedom. The maximum number of spatial modes that can be supported can be shown to be equal to the rank of the channel matrix, which in conventional MIMO with rich scattering is equal to the minimum between the number of transmit and receive antennas. 
The number of available spatial modes tends to decrease in LoS propagation conditions, mainly because the channel matrix may easily become rank-deficient. However, there exist configurations where the maximum number of spatial modes is available, even under LoS. 
For example, $M$ spatial streams are theoretically feasible in a MIMO system consisting of two parallel and confronted uniform linear arrays (ULAs) of $M$ elements, provided that the separation between consecutive antennas at the transmitter ($\Delta_T$) and at the receiver ($\Delta_R$) satisfy the \textit{Rayleigh spacing} criterion, that is \cite{Torkildson2011}
\begin{IEEEeqnarray}{c}
\Delta_T\Delta_R = \frac{D\lambda}{M}
\label{eq:Torkildson}
\end{IEEEeqnarray}
with $\lambda$ and $D$ denoting the wavelength and the distance between the arrays. 
The analysis is much more involved when the two ULAs are not confronted, although an upper bound on the capacity of LoS MIMO channel formed by two ULAs was presented in \cite{Do2021}. 
This bound was shown to be achievable 
by means of physically rotating the ULA depending on the signal to noise ratio (SNR). 

Zooming out to a more general context, the benefits obtained by using MIMO-surfaces and MIMO antenna arrays have recently been investigated in a number of contributions (see \cite{direnzo2023} and references therein). Holographic surface architectures are analyzed in detail in \cite{Hu18} under the assumption of sufficient large surface and employing the matched filter (MF) processing. The number of spatial modes was shown to be $\frac{2}{\lambda}$ modes per meter of deployed surface for one-dimensional terminal deployment, and $\frac{\pi}{\lambda^2}$ modes per squared meter for two and three dimensional terminal deployment. Similarly, analytical expressions for the link gain and the number of spatial streams when the communication is carried out between two holographic surfaces are provided in \cite{Dardari20}. It is worth pointing out that most of the previous literature relies on approximate models of the electromagnetic channel that essentially disregard the power differences between the different links and focus only on the phase effects of the propagation. This variability of the power has a significant effect in XL-MIMO configurations, since transmitting elements that are far away from the receiver are essentially irrelevant for the communication purposes.

\subsection{Use of polarization}
The use of orthogonally polarized antennas offers additional degrees of freedom in order to improve the spectral efficiency of the wireless channel. 
For example, by independently exciting three orthogonal dipoles one may theoretically support three different communication streams per array element, thus essentially multiplying by three the number of available spatial modes \cite{Andrews01}. 
The available spatial degrees of freedom in traditional (rich scattering) wireless communications is investigated in \cite{Poon05}, where only two polarizations are claimed to be effectively exploitable in the far field, even when working with antennas with multiple polarizations. The authors suggest that the third polarization might play a role in the \textit{near-} and \textit{intermediate-} field regions.


On the other hand, the use of dual-polarization antennas in XL-MIMO has recently been investigated in a number of works \cite{Chen21,Sena21,Wei23}. The analysis of the electromagnetic propagation phenomena reveals that inter-polarization interference may significantly affect the behavior of the resulting XL-MIMO channel. The attained spectral efficiency in the uplink of this type of polarized XL-MIMO architecture is addressed in \cite{Torres2020near}, from where one can conclude that the polarization mismatch can significantly degrade the supported spectral efficiency. In \cite{Ozdogan23} a mMIMO system with dual-polarized antennas is investigated, and closed-form expressions for the spectral efficiency are derived in the presence of different types of precoders. A Non-LoS communication scenario is assumed, where the equivalent channel can be modeled as a correlated Rayleigh fading with cross-polar correlation terms. Under this assumption, it is shown in \cite{Ozdogan23} that the use of two polarizations essentially allows to duplicate the supported spectral efficiency.


\subsection{Contributions}
This work is inspired by the polarization setting in \cite{Wei23} and considers the use of three orthogonally polarized antennas in the context of XL-MIMO. We depart from the conventional cross-polarization channel models employed in the literature \cite{Ozdogan23} and directly study the XL-MIMO channel that is generated by infinitesimal dipoles positioned in three orthogonal orientations \cite{Balanis05}. In the current setting, the electromagnetic channel can be characterized via the solution to the Helmholtz equation (i.e. the wave equation for the electric field). 
The main contributions of this paper can be summarized as:
\begin{itemize}


\item We characterize the achievable rate of the near-field XL-MIMO channel with full CSI at the transmitter side in the \textit{holographic regime}, whereby the number of array elements increases to infinity while the distance between consecutive elements converges to zero. Each element of the transmit receive array is assumed to consist of multiple orthogonal infinitesimal dipoles along the three spatial directions. As a by-product, we analyze the eigenvalues of the equivalent channel and establish the maximum number of achievable spatial degrees of freedom under different configurations. Results are derived for both ULAs and Uniform Planar Arrays (UPAs). 

\item Based on the above study, we establish and characterize the optimal aperture length that maximizes the spectral efficiency when the receiver is at distance $D$ of the center of the transmit array, elevation $\theta$ and is equipped with a single element of up to three orthogonal infinitesimal dipoles. Furthermore, we study under which circumstances the use of three orthogonal infinitesimal dipoles is needed at the transmitter and/or at the receiver sides.


\item We show that the requirements on the aperture length can be relaxed when adding multiple antennas at the receiver side, and optimizing their separation. Furthermore, we show that if the objective is to attain a percentage of the maximum spectral efficiency, the aperture length can be relaxed even more.

 \end{itemize}

Part of this work was previously published in \cite{agustin23}, which specifically addressed the case of an ULA transmitting to a single element receiver. In the present paper, we consider a Uniform Planar Array (UPA) and provide a more complete analysis for a higher number of receive antennas. 

The rest of the paper is organized as follows. Section~\ref{sec:system_model} presents the signal and channel models, introducing the transmit and receive antenna array configurations from the geometric point of view.  Section~\ref{sec:NumStreams} reviews the most important notions of MIMO communication theory that are needed in this work and formally establishes the defintion of spatial modes. 
Section~\ref{sec:channel_rank} 
provides asymptotic analytical expressions of the achievable rates in the \textit{holographic regime} when the transmitter is equipped with either an ULA or an UPA. Section~\ref{sec:results} provides a numerical analysis of the results and finally Section~\ref{sec:conclusion} concludes the paper. 

\section{System and Signal Model}
\label{sec:system_model}
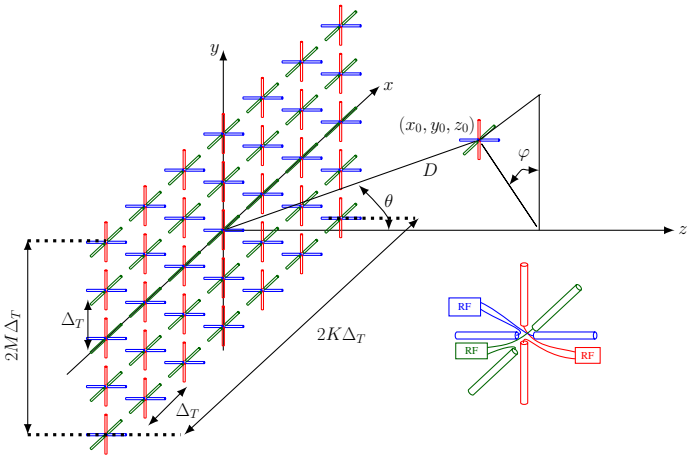
\begin{figure}[htbp]
\centerline{\begin{tikzpicture}[>={latex},scale=0.4,every node/.style={scale=0.4}]

\def\tripole at (#1,#2){
\node (A) [scale = 0.3, cylinder, shape border rotate=90, shape aspect=.5, draw = red, minimum height= 60, minimum width=1] at (#1+0,#2+0.35) {};
\node (A) [scale = 0.3, cylinder, shape border rotate=90, shape aspect=.5, draw = red, minimum height= 60, minimum width=1] at (#1+0,#2-0.35) {};
\node (A) [scale = 0.3,cylinder, shape border rotate=0, shape aspect=.5, draw = blue, minimum height= 60, minimum width=1] at (#1+0.35,#2+0) {};
\node (A) [scale = 0.3,cylinder, shape border rotate=0, shape aspect=.5, draw = blue, minimum height= 60, minimum width=1] at (#1-0.35,#2+0) {};
\node (A) [scale = 0.3,cylinder, rotate=43, shape aspect=.5, draw = black!60!green, minimum height= 60, minimum width=1] at (#1+0.35/1.25,#2+0.32/1.25) {};
\node (A) [scale = 0.3,cylinder, rotate=43, shape aspect=.5, draw = black!60!green, minimum height= 60, minimum width=1] at (#1-0.35/1.2,#2-0.33/1.2) {};
}

\def\tripoleGros at (#1,#2){
\node (A) [cylinder, shape border rotate=90, shape aspect=.5, draw = red, minimum height= 60, minimum width=1] at (#1+0,#2+1.3) {};
\node (A) [cylinder, shape border rotate=90, shape aspect=.5, draw = red, minimum height= 60, minimum width=1] at (#1+0,#2-1.3) {};
\node (A) [cylinder, shape border rotate=0, shape aspect=.5, draw = blue, minimum height= 60, minimum width=1] at (#1+1.3,#2+0) {};
\node (A) [cylinder, shape border rotate=0, shape aspect=.5, draw = blue, minimum height= 60, minimum width=1] at (#1-1.3,#2+0) {};
\node (A) [cylinder, rotate=43, shape aspect=.5, draw = black!60!green, minimum height= 60, minimum width=1] at (#1+1.3/1.25,#2+1.1/1.25) {};
\node (A) [cylinder, rotate=43, shape aspect=.5, draw = black!60!green, minimum height= 60, minimum width=1] at (#1-1.3/1.2,#2-1.5/1.2) {};
}

\draw [->] (0,0) node (v4) {} -- (15,0) node[right] {\LARGE $z$};
\draw [->] (0,-4) -- (0,6) node[left] {\LARGE $y$};
\draw [<-] (1.3*4,1.2*4) node[right] {\LARGE $x$} -- (-1.3*4,-1.2*4) ;

 \foreach \x in {-3,...,3}
    \foreach \y in {-2,...,2} 
     	\tripole at (1.3*\x,1.2*\x+1.6*\y);
    	 
 \tripole at (8.5,3);
 \node [left] at (8.5,3.5) {\LARGE $(x_0,y_0,z_0)$};
      
\draw (10.5,0) node (v3) {} -- (10.5,4.5) coordinate (v1) {} -- (v1) -- (8.5,3) node (v2) {} -- (v2) -- (v3) -- (v2) -- (v4);

\draw[<->] (4.5,1.5) .. controls (5,1) and (5.5,0.5) .. (5.5,0);
\draw[<->] (10.5,2) .. controls (9.9,2) and (10,2.15) .. (9.5,1.5);
\node at (5.5,1) {\LARGE $\theta$};
\node at (10,2.5)  {\LARGE $\varphi$};

\draw [<->] (-4.5,-4) -- (-4.5,-2.3);
\node at (-5,-3) {\LARGE $\Delta_T$};

\draw [<->] (-2.5,-6.5) -- (-1.2,-5.2);
\node [right] at (-1.7,-6) {\LARGE $\Delta_T$};

\draw [<->] (6.5,0.4) node (v5) {} -- (-1.3,-6.8) node (v6) {};
\node [right] at (3,-3.5) {\LARGE $2K\Delta_T$};
\draw [<->] (-6.5,-0.3) -- (-6.5,-6.8);
\node [rotate=90] at (-7,-3.5) {\LARGE $2M\Delta_T$};

\draw [dotted,very thick] (3.5,0.4) -- (v5);
\draw [dotted,very thick] (-6.5,-6.8) -- (v6);
\draw [dotted,very thick] (-6.5,-0.35) -- (-3.9,-0.35);
\node [right] at (6.5,2) {\LARGE $D$};

\tripoleGros at (10, -3.5); 

\begin{scope} [scale=1.5, shift={(-3.5,1.5)}]
\draw[red] (10.15,-3.62) .. controls (10.35,-4.1) and (10.9,-4.1) .. (11.3,-4.2);
\draw[red] (10.15,-4) .. controls (10.15,-3.7) and (10.5,-4.3) .. (11.3,-4.35);
\draw [blue] (9.99,-3.83) .. controls (10.34,-3.74) and (9.65,-3.54) .. (9.2,-3.25);
\draw [blue] (10.37,-3.84) .. controls (10.14,-3.81) and (9.65,-3.3) .. (9.2,-3.1);
\draw [black!60!green] (10.36,-3.68) .. controls (9.95,-4.05) and (9.7,-4.05) .. (9.35,-4.1);
\draw [black!60!green] (9.95,-4.15) .. controls (10.05,-4.05) and (9.8,-4.05) .. (9.35,-4.25);
\draw [blue] (9.2,-3) rectangle node {RF} (8.5,-3.4);
\draw [red] (11.85,-4.1) rectangle node {RF} (11.3,-4.45);
\draw [black!60!green] (8.65,-4) rectangle node {RF} (9.35,-4.35);

\end{scope}

\end{tikzpicture}}
\caption{Scenario configuration. The transmitter is based on a UPA consisting of $2K+1$ ULAs with $2M+1$ antenna elements each. In this paper, these elements consist of 2 dipoles (red and green in the figure) or 3 dipoles (fully polarized).}
\label{fig:Scenario}
\end{figure}

We consider a single-user XL-MIMO scenario as depicted in Fig. \ref{fig:Scenario}. The transmitter is an UPA centered on the $x$-$y$ plane, consisting of a matrix of $(2M+1) \times (2K+1)$ elements regularly placed with an interelement separation of $\Delta_T$ meters between consecutive elements\footnote{We assume an odd number of elements per raw and column in the UPA to ease the notation, although all the results are equally valid if the number of rows/columns of the UPA is even.}. Each element consists of up to three infinitesimal dipoles, each one assumed fed by a different radio frequency (RF) unit, so that multiplexing can also be achieved in the polarization domain. 

We will generally consider that the receiver may also consist of a multiple antenna array. In order to simplify the exposition, let us first focus on the channel seen by one of these receive antennas, which is assumed to be located at the position $(x_0,y_0,z_0)$ in canonical coordinates (we use the convention that $z_0>0$). Disregarding mutual coupling and other non-ideal effects, the received electric field at this point can be described as the solution to the Helmholtz wave equation, i.e. the Green function \cite{Poon05}. 
The equivalent channel matrix that is generated by the $(k,m)$th element of the UPA assuming infinitesimal dipole antennas is therefore given by \cite{Balanis05}
\begin{equation}
\mathbf{H}_{k,m}=\frac{\xi}{\lambda r_{k,m}}\exp\left(  -\mathrm{j}\frac{2\pi
}{\lambda}r_{k,m}\right)  \mathcal{H}_{k,m}\label{eq:channelModel}
\end{equation}
where $r_{k,m}$ is distance between the $(k,m)$th element of the transmit UPA 
and the receiver, $\lambda$ is the signal wavelength and $\xi$ is a complex 
constant that is proportional to the permittivity of the
medium\footnote{Indeed, if we take $\xi=\mathrm{j}\eta/2$ where $\eta$ is the
permittivity of the propagation medium, then $\mathbf{H}_{k,m}$ above is the
electric field caused by a unit electric current in each of the three spatial
directions. Here, $\xi$ contains also the conversion constant from
electric field into the corresponding digital signal at the receiver.}.
Finally, $\mathcal{H}_{k,m}$ is a $3\times3$  matrix given by
\[
\mathcal{H}_{k,m}=\alpha_{k,m}\mathbf{I}_{3}-\beta_{k,m}\frac{\mathbf{r}
_{k,m}{\mathbf{r}}_{k,m}^{H}}{\left\Vert {\mathbf{r}}_{k,m}\right\Vert ^{2}}
\]
where $\mathbf{r}_{k,m}$ denotes the position vector from the receiver to the $(k,m)$th element of the UPA and $r_{k,m}$ its norm, that is
\begin{align}
{\mathbf{r}}_{k,m} &  =\left[
\begin{array}
{c c c}
k\Delta_{T}-x_{0},
m\Delta_{T}-y_{0},
-z_{0}
\end{array}
\right] ^T \\
r_{k,m}^{2} &  =\left\Vert {\mathbf{r}}_{k,m}\right\Vert ^{2}=\left(
k\Delta_{T}-x_{0}\right)  ^{2}+\left(  m\Delta_{T}-y_{0}\right)  ^{2}%
+z_{0}^{2}.
\end{align}
Finally, the two coefficients $\alpha_{k,m}$ and $\beta_{k,m}$ are defined as
\begin{align*}
\alpha_{k,m} \!\! &   = \!1\!+\!\frac{ \mathrm{j}\lambda 2\pi r_{k,m}-\lambda^{2}
}{(2\pi r_{k,m})^{2}}, \,\,\,
\beta_{k,m} \!\! \!\! &  =\!1\!+3\frac{\mathrm{j}\lambda 2\pi r_{k,m}-\lambda^{2}
}{(2\pi r_{k,m})^{2}}.
\end{align*}

The above channel model assumes that the three dipoles at the receiver are
oriented along the $x$-$y$-$z$ axes of the system of coordinates. 
Otherwise, the channel in
(\ref{eq:channelModel}) is replaced with
\[
\mathbf{H}_{k,m}=\frac{\xi}{\lambda r_{k,m}}\exp\left(  -\mathrm{j}\frac{2\pi
}{\lambda}r_{k,m}\right)  \mathbf{Q}\mathcal{H}_{k,m}%
\]
where $\mathbf{Q}\in\mathbb{R}^{3\times3}$ is a certain rotation matrix that
accounts for the fact that the three receiving dipoles may not be placed along
the reference coordinate system. It can clearly be seen that this rotation
matrix does not really affect the capacity of the channel when three
polarizations are used, so that we will assume from now on that $\mathbf{Q=I}
_{3}$. Note, however, that this choice has geometric consequences when only two
polarizations are used at either side of the communications link, since it
basically implies that the corresponding dipoles are aligned with the system of coordinates.

When only a subset of polarizations are used at the transmitter/receiver, one
must select the corresponding entries of the above matrix, $\mathbf{H}_{k,m}$. In this paper, we
will assume that the first polarizations (according to the orientation of the
cartesian system of coordinates) are always used, so that we will consider the
corresponding channel model
\[
\mathbf{H}_{k,m}^{t_{\mathrm{pol}}\times r_{\mathrm{pol}}}=\left[
\mathbf{H}_{k,m}\right]  _{1:r_{\mathrm{pol}},1:t_{\mathrm{pol}}}
\]
where $t_{\mathrm{pol}}\ $and $r_{\mathrm{pol}}$ are the number of
polarizations used at the transmitter and the receiver respectively.

By observing the expression of the coefficients $\alpha_{k,m}$ and $\beta_{k,m}$,
one can realize that the channel matrix
can be decomposed as a sum of terms that decay as the inverse of
$r_{k,m}$, sometimes referred to as \textit{radiative terms}, plus terms decaying as
$O(r_{k,m}^{-2})$ and $O(r_{k,m}^{-3})$, sometimes referred to as \textit{reactive
terms}. From now on, we will fully disregard these two terms in the channel
definition, so that we will take
\begin{equation} \label{eq:approxchann}
\mathbf{H}_{k,m}\simeq\frac{\xi}{\lambda r_{k,m}}\exp\left(  -\mathrm{j}
\frac{2\pi}{\lambda}r_{k,m}\right)  \mathbf{P}_{k,m}^{\perp}
\end{equation}
where $\mathbf{P}_{k,m}^{\perp}$ is the orthogonal projection matrix onto the
nullspace of the propagation direction, that is
\[
\mathbf{P}_{k,m}^{\perp}=\mathbf{I}_{3}-\frac{\mathbf{r}_{k,m}\mathbf{r}%
_{k,m}^{H}}{\left\Vert \mathbf{r}_{k,m}\right\Vert ^{2}}.
\]
This approximation is made precise in Lemma \ref{lemma:radiative}, showing that it holds regardless of the geometry of the transmit array. 

\begin{lemma} \label{lemma:radiative}
Define the error matrix
\begin{align*}
\mathbf{E}_{k,m}^{t_{\mathrm{pol}}\times r_{\mathrm{pol}}}  & =\mathbf{H}
_{k,m}^{t_{\mathrm{pol}}\times r_{\mathrm{pol}}}\left(  \mathbf{H}
_{k,m}^{t_{\mathrm{pol}}\times r_{\mathrm{pol}}}\right)  ^{H}\\
& -\left\vert \frac{\xi}{\lambda}\right\vert ^{2}\frac{1}{r_{k,m}^{2}}\left[
\mathbf{P}_{k,m}^{\perp}\right]  _{1:r_{\mathrm{pol}},1:t_{\mathrm{pol}}
}\left[  \mathbf{P}_{k,m}^{\perp}\right]  _{1:r_{\mathrm{pol}}
,1:t_{\mathrm{pol}}}^{H}.
\end{align*}
The spectral norm of this matrix accepts the upper bound
\begin{align*}
\sup_{k,m}\left\Vert \mathbf{E}_{k,m}^{t_{\mathrm{pol}}\times r_{\mathrm{pol}
}}\right\Vert  &  \leq8\left\vert \frac{\xi}{\lambda}\right\vert ^{2}
\frac{\lambda}{2\pi d_{\inf}^{3}}\left(  1+\frac{\lambda}{2\pi d_{\inf}}\right)  \\
&  +32\left\vert \frac{\xi}{\lambda}\right\vert ^{2}\frac{\lambda^{2}}{\left(
2\pi\right)  ^{3}d_{\inf}^{4}}\left(  1+\left(  \frac{\lambda}{2\pi d_{\inf}
}\right)  ^{2}\right)  
\end{align*}
where $d_{\inf} = \inf_{k,m}r_{k,m}$ is the distance between the UPA and the receiver. When $t_{\mathrm{pol}}=3$ the upper bound simplfies to
\[
\sup_{k,m}\left\Vert \mathbf{E}_{k,m}^{t_{\mathrm{pol}}\times r_{\mathrm{pol}
}}\right\Vert \leq\left(  1+\frac{2}{\pi}\right)  \frac{\left\vert
\xi\right\vert ^{2}}{2\pi^{2}d_{\inf}^{4}}+\frac{\left\vert \lambda\xi\right\vert
^{2}}{2\pi^{5}d_{\inf}^{6}}.
\]
\end{lemma}
\begin{IEEEproof}
    See appendix \ref{sec:Ap_radiative}. 
\end{IEEEproof}
The above lemma formally justifies the use of the approximation in (\ref{eq:approxchann}). Noting that the Gramian of the channel matrix decays in spectral norm as the square of the distance between the UPA and the receiver, i.e. as $O(d_{\inf}^{-2})$, the above lemma states that the error term decays at least as $O(d_{\inf}^{-3})$ and that the convergence rate is even faster when three polarizations are used at the transmitter. Furthermore, the result holds uniformly in the transmit antenna index, so that averages of the Gramians can also be approximated up to the same order of magnitude of the error terms. In particular, the approximation will also hold in the \textit{holographic asymptotic regime} (i.e. a large number of transmit antennas with an asymptotically small spacing). 

In order to provide an expression for the received signal, let us consider the 
channel matrix that is obtained by stacking, side by side, the $r_\mathrm{pol} \times
t_\mathrm{pol}$ channel matrices from each of the $(2K+1)(2M+1)$ elements of the UPA. The 
result is an $r_\mathrm{pol} \times (2K+1)(2M+1)t_\mathrm{pol}$ matrix given by, 
\begin{equation} \label{eq:finalChannel}
\mathbf{H}^{t_\mathrm{pol} \times r_\mathrm{pol}} \in  \mathbb{C}^{r_\mathrm{pol} \times (2K+1)(2M+1)t_\mathrm{pol}}.
\end{equation}
Let $\mathbf{x}$ define the 
$r_\mathrm{pol}$-dimensional vector containing the signals intended for the receiver, which
are assumed to be zero mean and unit power. Let $\mathbf{F}$ denote a $(2M+1)(2K+1)t_{\mathrm{pol}
}\times r_{\mathrm{pol}}$ precoding matrix that transforms the
$r_{\mathrm{pol}}$ signals intended for the receiver into a set of
$(2M+1)(2K+1)t_{\mathrm{pol}}$ signals that are fed to the different transmit
dipoles. This precoder is designed assuming perfect CSI at the transmitter with a total transmit power constraint equal to
$\mathrm{tr}(\mathbf{FF}^{H})\!\!=\!\!P$. The received signal is expressed as
\[
\mathbf{y}=\mathbf{H}^{t_{\mathrm{pol}}\times r_{\mathrm{pol}}}\mathbf{Fx}+\mathbf{n}
\]
where $\mathbf{n}$ is a circularly symmetric complex Gaussian distributed noise with zero 
mean and power $\sigma^2$. 
\begin{remark}
\label{rem:expand_signal_model}
    The above signal model has been obtained for the specific case where the receiver consists of a single spatial element with three orthogonal infinitesimal dipoles. The model can however be trivially generalized to the case where the receiver consists of an antenna array, by simply stacking the corresponding channel matrices in (\ref{eq:channelModel}) on top of one another. 
\end{remark}

\section{Effective number of DoF} \label{sec:NumStreams}

Assuming Gaussian signalling, we can express the achievable rate of the above system (in bits/s/Hz) as
\[
\mathrm{C}=\log_2\det\left(  \mathbf{I}_{r_{\mathrm{pol}}}\mathbf{+}\frac
{1}{\sigma^{2}}\mathbf{H}^{t_{\mathrm{pol}}\times r_{\mathrm{pol}}}
\mathbf{FF}^{H}\left(  \mathbf{H}^{t_{\mathrm{pol}}\times r_{\mathrm{pol}}
}\right)  ^{H}\right).
\]
Now, it is well known that the optimum
precoder $\mathbf{F}$ is designed as follows, \cite{Tse05, Palomar05}. Consider the singular value
decomposition 
$\mathbf{H}^{t_{\mathrm{pol}}\times r_{\mathrm{pol}}
}=\mathbf{U\Gamma}^{1/2}\mathbf{V}^{H}$ where $\mathbf{U}\in\mathbb{C}
^{r_{\mathrm{pol}}\times r_{\mathrm{pol}}}$ is the matrix of left singular
vectors, $\mathbf{V}\in\mathbb{C}^{(2M+1)t_{\mathrm{pol}}\times
r_{\mathrm{pol}}}$ is the matrix of right singular vectors, and
$\mathbf{\Gamma}^{1/2}\mathbf{\in}\mathbb{C}^{r_{\mathrm{pol}}\times
r_{\mathrm{pol}}}$ is a diagonal matrix containing the singular values of
$\mathbf{H}^{t_{\mathrm{pol}}\times r_{\mathrm{pol}}}$, denoted by 
$\sqrt{\gamma_1} \geq \ldots \geq \sqrt{\gamma_{r_\mathrm{pol}}}$. The optimum precoding matrix takes the form
$\mathbf{F}=\mathbf{V}\mathcal{P}^{1/2}$ where $\mathcal{P}^{1/2}$ is a
diagonal matrix with non-negative coefficients $\sqrt{p_{1}},\ldots,\sqrt{p_{r_{\mathrm{pol}}}}$ that are chosen according to the waterfilling equation
\[
p_{i}=\left[  \frac{1}{\mu}-\frac{\sigma^{2}}{\gamma_{i}}\right]^{+}
\]
where $[\cdot]^{+}\!\!=\!\!\max(\cdot,0)$. The water level, $\mu$, is obtained by 
\[
P=\sum_{j=1}^{r_{\mathrm{pol}}}p_{j}=\sum_{i=1}^{r_\mathrm{pol}}\left[  \frac
{1}{\mu}-\frac{\sigma^{2}}{\gamma_{i}}\right]  ^{+}.
\]
The number of spatial degrees of freedom (DoF), or maximum number of spatial transmission modes is defined by 
\begin{IEEEeqnarray}{c}
\bar{\nu}^{t_\mathrm{pol} \times r_\mathrm{pol}} = \lim_{\frac{P}{\sigma^2}\rightarrow\infty}\frac{ \mathrm{C}}{\log_2(P/\sigma^2)}. 
\label{eq:dof}
\end{IEEEeqnarray}
It can easily be seen that when the  
($P/\sigma^2$) is sufficiently high 
all the power coefficients are positive, so that $\bar{\nu}^{t_\mathrm{pol} \times r_\mathrm{pol}} $ coincides in practice with the 
number of nonzero singular values of the channel matrix in (\ref{eq:finalChannel}).
We will alternatively denote as ${\nu}^{t_\mathrm{pol} \times r_\mathrm{pol}}$ the corresponding
quantity for finite values of $P/\sigma^2$, which will be referred to as the effective number
of degrees of freedom.

We can get some insights about the eigenvalues exploring the structure of Gramian $\mathcal{W}^{t_{\mathrm{pol}}\times r_{\mathrm{pol}}} = \mathbf{H}
^{t_{\mathrm{pol}}\times r_{\mathrm{pol}}}\left(  \mathbf{H}^{t_{\mathrm{pol}}\times 
r_{\mathrm{pol}}}\right)  ^{H}$. Taking into account Lemma \ref{lemma:radiative} and disregarding all the reactive terms, the Gramian is given by 
\begin{equation} \label{eq:defW}
\!\!\!{\mathcal{W}}^{t_{\mathrm{pol}}\times r_{\mathrm{pol}}} \!\! =\!\!\sum_{m,k}\!\frac{\vert \xi /\lambda \vert^2}{r_{k,m}^{2}} 
\!\!\left[  \mathbf{P}_{k,m}^{\perp}\right]
_{1:r_\mathrm{pol},1:t_{\mathrm{pol}}} \!\!\left[  \mathbf{P}_{k,m}^{\perp}\right]
_{1:r_\mathrm{pol},1:t_{\mathrm{pol}}}^{H} 
\end{equation}
where the sum is over all antenna indices. In the following section, we will study the behavior of this matrix when the number of antenna elements grows without bound whereas  the separation converges to zero at the same rate. 
Now, it can easily seen that as the number of antennas grow, the eigenvalues of the Gramian ${\mathcal{W}}^{t_{\mathrm{pol}}\times r_{\mathrm{pol}}}$ increase in magnitude, so that the overall capacity increases 
in magnitude as well. This power boosting effect is a well known consequence of the fact that the
availability of CSI at the transmitter allows to focus the energy in a certain spatial region with increasingly high accuracy as the number of antennas grows. 
Thus, it makes sense to scale down the total transmit power as the number of transmit elements grows without bound, so that the total system capacity stays bounded. 
To that end, we will assume that we fix the total transmitted power as  
\[
P=\frac{\bar{P}}{\left(  2M+1\right)\left(  2K+1\right)   t_{\mathrm{pol}}}
\]
where $\bar{P}$ is a constant independent of the number of transmit antennas. 

We define the \emph{reference} SNR as the signal to noise ratio that is achieved at the receiver when the transmit array consists of a single antenna ($K=M=0$) and three polarizations are used at either side of the communications link. It can readily be seen that this reference SNR takes the form
\begin{equation} \label{eq:SNR_rx_def}
\mathsf{SNR_{0}} = \frac{\bar{P}}{\sigma^{2}t_{\mathrm{pol}}}\left\vert \frac{\xi}{\lambda}\right\vert^{2} \frac{1}{D^2}
\end{equation}
where $D$ is the distance between the receiver and the center of the array. 

Let us now consider a normalized version of the Gramian in (\ref{eq:defW}), that is
\begin{equation} \label{eq:norm_W}
\!\!\!\!    \mathbf{W}^{t_\mathrm{pol}\times r_\mathrm{pol}} = \sum_{k,m} \frac{\left[  \mathbf{P}_{k,m}^{\perp}\right]
_{1:r_\mathrm{pol},1:t_{\mathrm{pol}}} \left[  \mathbf{P}_{k,m}^{\perp}\right]
_{1:r_\mathrm{pol},1:t_{\mathrm{pol}}}^{H}}{(2M+1)(2K+1) (r_{k,m}^2/D^2)}
\end{equation}
and let $\rho_i$, $i=1,\ldots,r_\mathrm{pol}$ denote its eigenvalues, sorted in descending order of magnitude. With this definition, the spectral efficiency of the system is given by
\begin{equation} \label{eq:capacity_normalized}
\mathrm{C} =\sum_{i=1}^{r_\mathrm{pol}}\log_2\left(  1+{\rho_{i}}\tilde
{p}_{i}\right) 
\end{equation}
where now $\rho_{i}$ are the eigenvalues of the Gramian, $\tilde{p}_{i}$ are now defined as $\tilde{p}_{i}=[ {\vartheta}^{-1}-{\rho_{i}}^{-1}] ^{+}$
and 
$\vartheta$ is fixed by forcing $\mathsf{SNR_{0}}=\sum_{j=1}^{r_{\mathrm{pol}}}\tilde{p}_{i}$. 
Now, based on these definitions, we can characterize the number of active streams 
that are active in certain geographical areas. In fact, when $r_\mathrm{pol} = 3$, 
we can establish that the total number of active streams is given by 
\[
n^+ = \left\{
\begin{array}
[c]{cc}
1 & \mathsf{SNR_{0}} < \mathsf{SNR}_\mathrm{th}^{(1)} \\
2 &  \mathsf{SNR}_\mathrm{th}^{(1)} \leq \mathsf{SNR_{0}} < \mathsf{SNR}_\mathrm{th}^{(2)}\\
3 &  \mathsf{SNR}_\mathrm{th}^{(2)}\leq \mathsf{SNR_{0}} .
\end{array}
\right. 
\]
where we have introduced the thresholds $\mathsf{SNR}_\mathrm{th}^{(1)} = \rho_2^{-1} - \rho_1^{-1}$ and $\mathsf{SNR}_\mathrm{th}^{(2)} = 2\rho_3^{-1} - \rho_1^{-1} - \rho_2^{-1}$,
which only depend on the eigenvalues of $\mathbf{W}^{t_{\mathrm{pol}}\times r_{\mathrm{pol}}}$. 
This way, the characterization of the eigenvalues of the Gramian $\mathbf{W}^{t_{\mathrm{pol}}\times r_{\mathrm{pol}}}$ in (\ref{eq:defW}) will directly establish the number of spatial modes that can be exploited at each geographical point of the spatial domain, depending on the reference $\mathsf{SNR_{0}}$ that can be obtained at that point.

\section{Asymptotic behavior (holographic regime)}
\label{sec:channel_rank}
This section studies the behavior of the channel in the holographic regime, that is
when the total number of array elements converges to infinity while the distance
between consecutive elements converges to zero ($\Delta_T  \rightarrow  0$). We will first
focus for simplicity on the ULA case (namely $K = 0$ while $M \rightarrow \infty$) in Section~ref{sec:ULA}. Afterwards, in Section~\ref{sec:UPA}, we turn our
attention to the UPA case where both $K$ and $M$ tend to infinity at the same
rate so that $K\Delta_T  \rightarrow  L_x$ and $M\Delta_T \rightarrow L_y$. 


\subsection{Transmitter with ULA and receiver with a single element}
\label{sec:ULA}
Following the description in Fig.~\ref{fig:Scenario}, we consider here a simple case where the transmitter is equipped with a single ULA of $2M+1$ elements placed along the $y$-axis, so that  $K=0$, while the receiver only has one receiving element that is located on the $y$-$z$ plane, namely at the location $(0,y_0,z_0)$. This configuration is slightly more general than the one that was studied in \cite{agustin23} (receiver on the broadside of the array, i.e. $x_0 = y_0 = 0$), and results can be generalized with little effort to the case $x_0 \neq 0$. We will see later in Section~\ref{sec:UPA} that the ULA case with $x_0 \neq 0$ can be obtained as a particularization of the UPA study, so we omit it here for the sake of simplicity. 

In order to introduce the results in this section, it will prove useful to express the position of
the receiver in polar coordinates. We recall that $D$ denotes the total distance between the center
of the transmit array and the receiver, so that we can express $y_0 = D\sin\theta$ and $z_0 = D\cos\theta$, where $\theta \in (-\pi,\pi)$ is the elevation angle in polar coordinates. We will study 
the behavior of the matrix in (\ref{eq:norm_W}) as $M \rightarrow \infty$, $\Delta_T \rightarrow 0$ while $M\Delta_T \rightarrow L$ for some positive $L$ that can be seen as half of the total length of the ULA. 
This asymptotic regime represents a \textit{holographic} multi-antenna configuration, according to which
an infinite number of antennas are placed infinitely close to one another in a segment of total
length $2L$ meters.

In order to present the result, we will first define three quantities that will be of outmost
importance for the consequent analysis, namely $\psi_{2}$, $\psi_{3}$ and $\psi_{4}$. In fact,
it can easily be seen that $D^k\psi_k$ is a function of the elevation $\theta$ and the ratio between the size of the ULA and the distance, which will be denoted as $\varrho=L/D$. This fact will be exploited later in order to show that there exists an optimum array length to maximize
the spectral efficiency. The first quantity, $\psi_{2}$ takes the form 

\begin{equation}  \label{eq:DefPsi2}
 \!\!\!\!   \psi_2 = \frac{1}{2D^2 \varrho  \cos \theta} \left[ \arctan \frac{\varrho-\sin\theta}{\cos\theta} \! + \! \arctan \frac{\varrho + \sin\theta}{\cos{\theta}} \right] 
\end{equation}
where we point out that the quantity between brackets is the angle of view of the ULA
from the receiver (see also the next subsection). 
Regarding the quantities $\psi_{3}$ and $\psi_{4}$, they are defined as
\begin{align*}
\psi_{3}  &  =-\frac{1}{D^3}\frac{\sin\theta}{\left(  1+\varrho^2\right)  ^{2}-\left(
2\varrho\sin\theta\right)^{2}}\\
\psi_{4}  &  =\frac{1}{2D^4}\frac{1}{  \cos^{2}\theta}  \left[
\frac{\left(  1+\varrho^2\right)  -2\sin^{2}\theta}{\left(  1+\varrho^2\right)  ^{2}-\left(  2\varrho\sin\theta\right)  ^{2}}+ D^2\psi_{2}\right].
\end{align*}
With these definitions, we are now in the position to formulate the first asymptotic result,
which focuses on the case $(t_\mathrm{pol}\times r_\mathrm{pol})=(3,3)$. The corresponding behavior for lower $r_\mathrm{pol}$ 
follows by selecting the $r_\mathrm{pol} \times r_\mathrm{pol}$ upper left corner of the
corresponding matrices. 

\begin{proposition} \label{prop:ULA3x3}
Assume $K=0$ (transmit ULA) and $t_\mathrm{pol} = r_\mathrm{pol} = 3$. As $M\rightarrow\infty$ while $\Delta_{T}\rightarrow0$ and $M\Delta_{T}\rightarrow L$, the matrix $\mathbf{W}^{3\times3}$ converges to $\overline{\mathbf{W}}^{3\times3}$, which is given by
\[
\overline{\mathbf{W}}^{3\times3}= D^2 \left[
\begin{array}
[c]{ccc}%
\psi_{2} & 0 & 0\\
0 & \psi_{4}D^{2}\cos^{2}\theta & \psi_{3}D\cos\theta\\
0 & \psi_{3}D\cos\theta & \psi_{2}-\psi_{4}D^{2}\cos^{2}\theta
\end{array}
\right].
\]
\end{proposition}
\begin{IEEEproof}
The result follows from a direct application of the definition of Riemann integral, see Appendix~\ref{sec:proofsPropsULA}. 
\end{IEEEproof}
From the convergence of the Gramian $\mathbf{W}^{3\times 3}$ towards $\overline{\mathbf{W}}^{3\times 3}$ it directly follows that the achievable rate also converges to a fixed quantity, which is still given by (\ref{eq:capacity_normalized}) but simply replacing the eigenvalues of $\mathbf{W}^{3\times 3}$ with those of $\overline{\mathbf{W}}^{3\times 3}$. Hence, the above results offers a very convenient way of analyzing the behavior of the system in terms of its asymptotic achievable rate. In particular, it can readily be seen that the asymptotic Gramian $\overline{\mathbf{W}}^{3\times3}$ depends only on  $(\varrho, \theta)$. Now, when the transmit power increases without bound, the spectral efficiency of the {holographic} system at high SNR will behave as 
\[
\overline{\mathrm{C}}\left(\varrho,\theta\right) = \log_2 \det \left({\mathsf{SNR}}_{0} \  \overline{\mathbf{W}}^{3\times3} \right) + o(\log \mathsf{SNR}_{0}).
\]
It will be reasoned in Section~\ref{sec:results} that the first term is maximum at a finite non-zero $\varrho$ for every possible elevation $\theta$. 

 A similar asymptotic behavior can be described in the case $\left(t_\mathrm{pol}\times r_\mathrm{pol}\right)=(2,3)$. 
 In the 
 following proposition, we use the short hand notation $D_s = D \sin\theta$ and $D_c = D \cos\theta$ to emphasize the dependence on $D$ (note that these quantities are in fact
 equal to $y_0$ and $z_0$ respectively).
\begin{proposition} \label{prop:ULA2x3}
    Assume $K=0$ (transmit ULA), $t_\mathrm{pol} =2$ and $r_\mathrm{pol} = 3$. As $M\rightarrow\infty$ while $\Delta_{T}\rightarrow0$ and $M\Delta_{T}\rightarrow L$, the matrix $\mathbf{W}^{2\times3}$ converges to $\overline{\mathbf{W}}^{2\times3}$ given by
    \[
    \overline{\mathbf{W}}^{2 \times 3}= D^2
    \left[
    \begin{array}
    [c]{ccc}
    \psi_{2} & 0 & 0\\
    0 &   D_c^{4}\psi_{6} & D_c^{3}\psi_{5}\\
    0 & D_c^{3}\psi_{5} &  D_c^{2}\psi_{4}- D_c^{4}\psi_{6}
    \end{array}
    \right]  
    \]
where $\psi_{2}$, $\psi_{3}$ and $\psi_{4}$ are defined in Proposition
\ref{prop:ULA3x3} and where we have introduced the additional quantities
\begin{align}
\psi_{5}  &  =-\frac{1}{D^5}\frac{\left(  1+\varrho^{2}\right)  \sin{\theta}}{\left(  \left(
1+\varrho^{2}\right)  ^{2}-\left(  2\varrho\sin\theta\right)  ^{2}\right)  ^{2}} \label{eq:defpsi5}\\
\psi_{6}  &  =\frac{1}{4D^6}\frac{1}{\cos^2\theta}
\frac{\left(  1+\varrho^{2}\right)  ^{2}-4\sin^2\theta}{\left(  \left(
1+\varrho^{2}\right)  ^{2}-\left(  2\varrho\sin\theta\right)  ^{2}\right)  ^{2}} \nonumber \\
&  +\frac{3}{8D^6}\frac{1}{ \cos^{4}\theta}\left[  \frac{\left(
1+\varrho^{2}\right)  -2\sin^{2}\theta}{\left(  1+\varrho^{2}\right)
^{2}-\left(  2\varrho\sin\theta\right)  ^{2}}+D^2\psi_{2}\right]  . \label{eq:defpsi6}
\end{align}

\end{proposition}
\begin{IEEEproof}
    See Appendix \ref{sec:proofsPropsULA}. 
\end{IEEEproof}

Here similar conclusions can be drawn as in the case $t_{pol} = 3$, i.e. in the \textit{holographic} regime
 matrix $ \overline{\mathbf{W}}^{2 \times 3}$ becomes a function of $(\varrho,\theta)$. 
 In Section~\ref{sec:results} we will use the above expressions to study the effect of using multiple polarizations in a practical setting. 

\subsection{Transmitter with UPA and receiver with a single element}
\label{sec:UPA}

We study here the behavior of the scenario where the transmitter consists
of a UPA on the $x$-$y$ plane and the receiver consists of a single 
element located in the position $(x_0,y_0,z_0)$, see Fig.~\ref{fig:Scenario}. 
The total dimensions of the UPA can be expressed as $2L_x \times 2L_y$ square meters. In this asymptotic regime, the entries of the Gramian matrix $\mathbf{W}^{t_\mathrm{pol} \times r_\mathrm{pol}}$ in (\ref{eq:norm_W}) also converge to a constant independent of the number of transmit antennas. 

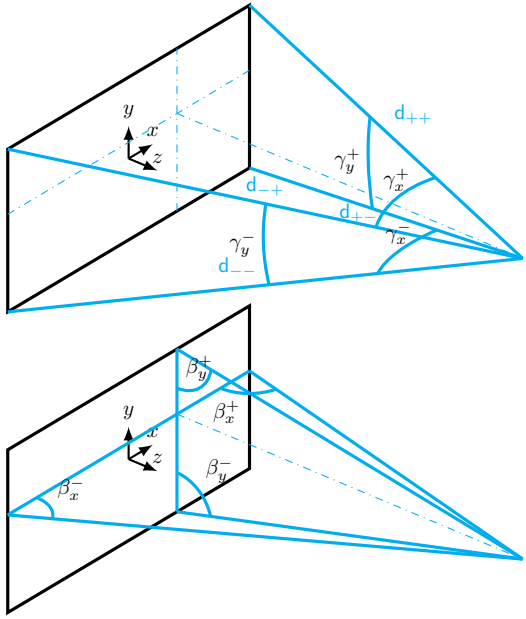
\begin{figure}[htbp]
\centerline{\tdplotsetmaincoords{120}{-40}
\begin{tikzpicture}[>={latex},scale=0.5,every node/.style={scale=0.8},tdplot_main_coords]

\pgfmathsetmacro{\Lx}{5}
  \pgfmathsetmacro{\Ly}{2.5}
  \pgfmathsetmacro{\Rx}{2}
  \pgfmathsetmacro{\Ry}{.5}
  \pgfmathsetmacro{\Rz}{12}
  
\pgfmathsetmacro{\r1}{2}
\pgfmathsetmacro{\r2}{1.75}
\pgfmathsetmacro{\r3}{1.5}
\pgfmathsetmacro{\r4}{1}

\begin{scope}
\draw[thick,->] (0,0,0) -- (1,0,0) node[anchor=south]{$z$};
\draw[thick,->] (0,0,0) -- (0,-1,0) node[anchor=south ]{$x$};
\draw[thick,->] (0,0,0) -- (0,0,1) node[anchor=south]{$y$};
\coordinate (pp) at (0,-\Lx,\Ly);
\coordinate (pm) at (0,-\Lx, -\Ly);
\coordinate (mp) at (0,\Lx,\Ly);
\coordinate (mm) at (0,\Lx,-\Ly);
\coordinate (vv) at (0,-\Rx,\Ry);
\coordinate (RX) at (\Rz,-\Rx,\Ry);
\draw [very thick] (pp)--(pm)--(mm)--(mp)--cycle;
\draw [cyan,dashdotted] (vv)--(RX);
\draw [cyan,dashdotted] (0,-\Rx,-\Ly) -- (0,-\Rx,\Ly);
\draw [cyan,dashdotted] (0,-\Lx,\Ry) -- (0,\Lx,\Ry);
\draw [cyan,very thick] (0,-\Lx,\Ly) --node[above right] {$\mathsf{d}_{++}$} (RX);
\draw [cyan,very thick] (0,-\Lx,-\Ly) --node[left] {$\mathsf{d}_{+-}$}  (RX);
\draw [cyan,very thick] (0,\Lx,-\Ly) --node[above left] {$\mathsf{d}_{--}$} (RX);
\draw [cyan,very thick] (0,\Lx,\Ly) --node[above] {$\mathsf{d}_{-+}$} (RX);
\tdplotdefinepoints(\Rz,-\Rx,\Ry)(0,\Lx,\Ly)(0,-\Lx,\Ly)
\tdplotdrawpolytopearc[very thick, cyan]{4}{anchor=south}{$\gamma_x^{+}$}
\tdplotdefinepoints(\Rz,-\Rx,\Ry)(0,-\Lx,-\Ly)(0,\Lx,-\Ly)
\tdplotdrawpolytopearc[very thick, cyan]{4}{anchor=south}{$\gamma_x^{-}$}
\tdplotdefinepoints(\Rz,-\Rx,\Ry)(0,-\Lx,-\Ly)(0,-\Lx,\Ly)
\tdplotdrawpolytopearc[very thick, cyan]{7}{anchor=east}{$\gamma_y^{+}$}
\tdplotdefinepoints(\Rz,-\Rx,\Ry)(0,\Lx,\Ly)(0,\Lx,-\Ly)
\tdplotdrawpolytopearc[very thick, cyan]{7}{anchor=east}{$\gamma_y^{-}$}
\end{scope}

\begin{scope}[yshift = - 8 cm]
 \draw[thick,->] (0,0,0) -- (1,0,0) node[anchor=south]{$z$};
 \draw[thick,->] (0,0,0) -- (0,-1,0) node[anchor=south ]{$x$};
 \draw[thick,->] (0,0,0) -- (0,0,1) node[anchor=south]{$y$};
\coordinate (pp) at (0,-\Lx,\Ly);
\coordinate (pm) at (0,-\Lx, -\Ly);
\coordinate (mp) at (0,\Lx,\Ly);
\coordinate (mm) at (0,\Lx,-\Ly);
\coordinate (vv) at (0,-\Rx,\Ry);
\coordinate (RX) at (\Rz,-\Rx,\Ry);
\draw [very thick] (pp)--(pm)--(mm)--(mp)--cycle;
\draw [cyan,dashdotted] (vv)--(RX);
\draw [cyan,very thick] (0,-\Rx,-\Ly) -- (0,-\Rx,\Ly);
\draw [cyan,very thick] (0,-\Lx,\Ry) -- (0,\Lx,\Ry);
\draw [cyan,very thick] (0,-\Rx,\Ly) -- (RX);
\draw [cyan,very thick] (0,-\Rx,-\Ly) -- (RX);
\draw [cyan,very thick] (0,\Lx,\Ry) -- (RX);
\draw [cyan,very thick] (0,-\Lx,\Ry) -- (RX);
\tdplotdefinepoints(0,-\Rx,\Ly)(\Rz,-\Rx,\Ry)(0,-\Rx,-\Ly)
 \tdplotdrawpolytopearc[very thick, cyan]{1.2}{anchor=south }{$\beta_y^{+}$}
 \tdplotdefinepoints(0,-\Rx,-\Ly)(\Rz,-\Rx,\Ry)(0,-\Rx,\Ly)
 \tdplotdrawpolytopearc[very thick, cyan]{1.2}{anchor=south west}{$\beta_y^{-}$}
  \tdplotdefinepoints(0,-\Lx,\Ry)(\Rz,-\Rx,\Ry)(0,\Lx,\Ry)
 \tdplotdrawpolytopearc[very thick, cyan]{1.2}{anchor=north east}{$\beta_x^{+}$}
  \tdplotdefinepoints(0,\Lx,\Ry)(\Rz,-\Rx,\Ry)(0,-\Lx,\Ry)
 \tdplotdrawpolytopearc[very thick, cyan]{1.2}{anchor=south west}{$\beta_x^{-}$}
 
%
\end{scope}

\end{tikzpicture}}
\caption{Geometry of the UPA scenario with the definition of the main angles at the transmitter(bottom) and distances and angles from receiver (top).}
\label{fig:Angles}
\end{figure}

In order to present these results,  Fig.~\ref{fig:Angles} introduces some geometric quantities (distances and angles) that come into play in the asymptotic description of the channel matrix. More specifically, we define the four distances between receiver and the four vertices of the UPA as $\mathsf{d}_{++}$ (upper right), $\mathsf{d}_{+-}$ (lower right), $\mathsf{d}_{-+}$ (upper left) and $\mathsf{d}_{--}$ (lower left), see further the upper diagram in Fig.~\ref{fig:Angles}. Still in this diagram, we define the four angles with which the receiver sees the four edges of the UPA as $\gamma_x^-$ (lower horizontal), $\gamma_x^+$ (upper horizontal), $\gamma_y^-$ (left vertical), $\gamma_y^+$ (right vertical). For example, we have
\begin{align*}
\!\!\!\!\!\! \gamma_y^{-}=\!\arctan
\frac{L_{y}-y_{0}}{\sqrt{\left(  L_x\!-\!x_{0}\right)  ^{2}\!+\!z_{0}^{2}}}
\!+\!\arctan  \frac{L_{y}\!+\!y_{0}}{\sqrt{\left(  L_x\!-\!x_{0}\right)  ^{2}
\!+\!z_{0}^{2}}}
\end{align*}
and a similar definition holds for the other angles. Now, moving forward to the lower diagram in Fig.~\ref{fig:Angles}, we consider the four angles that are formed by the UPA plane ($z=0$) with the four triangular sides of the tetrahedron in the upper plot. More specifically, $\beta_x^+$ (resp. $\beta_x^-$) is the angle between the UPA and the right (resp. left) triangular side of the tetrahedron. Similarly, $\beta_y^+$ and $\beta_y^-$ are the angles between the UPA plane and the upper and lower triangular sides of the tetrahedron respectively, see lower plot in Fig.~\ref{fig:Angles}. In particular, we can define 
\begin{align*}
\cos\mathcal{\beta}_{x}^{+} = \frac{L_{x} - x_{0}}{\sqrt{\left(  L_{x} - x_{0}\right)  ^{2} + z_{0}^{2}}}, 
\cos\mathcal{\beta}_{y}^{+} = \frac{L_{y} - y_{0}}{\sqrt{\left(  L_{y} - y_{0}\right)  ^{2} + z_{0}^{2}}}
\end{align*}
and similar definitions hold for the rest of the angles. 

The following proposition establishes the asymptotic behavior of
the Gramian matrix in (\ref{eq:norm_W}) in the holographic regime when $t_\mathrm{pol} = r_\mathrm{pol}= 3$.
\begin{proposition} \label{prop:UPA3x3}
Assume $t_\mathrm{pol} = r_\mathrm{pol} = 3$. As $M,K\rightarrow\infty$ while $\Delta_{T}\rightarrow0$ so that $M\Delta_{T}\rightarrow L_y$ and $K\Delta_{T}\rightarrow L_x$, the matrix $\mathbf{W}^{3\times3}$ converges to $\overline{\mathbf{W}}^{3\times3}$, which is given by equation (\ref{eq:AsymGrammUPA3x3}) at the top of the next page, with the short-hand notations $s_x^+ =\sin \beta_x^+, c_x^+ =\cos \beta_x^+,s_x^- =\sin \beta_x^-, c_x^- =\cos \beta_x^- $ and the equivalent definitions for $s_y^\pm, c_y^\pm$ changing $\beta_x^\pm$ for $\beta_y^\pm$ (see Fig. \ref{fig:Angles}).
Additionally, we have introduced the definition\footnote{This function can be evaluated as a single integral, see further Appendix~\ref{sec:UPA3x32x3}.}
\[
\Phi_2 = \frac{1}{4 L_x L_y} \int_{-L_x}^{L_x}\int_{-L_y}^{L_y} \frac{1}{(x-x_0)^2+(y-y_0)^2+z_0^2}dx dy.
\]
\end{proposition}
\begin{IEEEproof}
    See Appendix \ref{sec:UPA3x32x3}.
\end{IEEEproof}
It is interesting to observe that the asymptotic Gramian of the channel matrix $\overline{\mathbf{W}}^{3 \times 3}$ only depends on the angles of the tetrahedron in Fig.~\ref{fig:Angles} and not on the absolute values of the distances of the different edges. Indeed, this follows trivially from the analytical expressions in (\ref{eq:AsymGrammUPA3x3}) together with the fact $D^2\Phi_2$ can alternatively be expressed as a function of the angles in the scenario by simply applying a change of variables to polar coordinates in the definition of $\Phi_2$. With this, one can conclude that $\overline{\mathbf{W}}^{3 \times 3}$ can be determined by simply considering the relative position of the receiver with respect to the UPA, i.e. position of receiver and aspect ratio of the UPA.

\begin{figure*}[t]
    \centering
    \normalsize
    \begin{equation}   \label{eq:AsymGrammUPA3x3}
    \!\!\!\overline{\mathbf{W}}^{3\times3}=\Phi_{2}D^2\left[
\begin{array}
[c]{ccc}%
\frac{1}{2} & 0 & 0\\
0 & \frac{1}{2} & 0\\
0 & 0 & 1
\end{array}
\right] \! +\frac{D^2}{8L_{x}L_{y}}\left[
\begin{array}
[c]{ccc}%
c_{x}^{+}\gamma_{y}^{+}+c_{x}^{-}\gamma_{y}^{-} & \log\frac{\mathsf{d}_{--}\mathsf{d}_{++}}{\mathsf{d}_{+-}\mathsf{d}_{-+}} & -(  s_{x}^{+}\gamma_{y}^{+}-s_{x}^{-}\gamma_{y}^{-}) \\
\log\frac{\mathsf{d}_{--}\mathsf{d}_{++}}{\mathsf{d}_{+-}\mathsf{d}_{-+}}
& c_{y}^{+}\gamma_{x}^{+}+c_{y}^{-}\gamma_{x}^{-} & - (  s_{y}^{+}%
\gamma_{x}^{+}-s_{y}^{-}\gamma_{x}^{-} ) \\
-(  s_{x}^{+}\gamma_{y}^{+}-s_{x}^{-}\gamma_{y}^{-}) & -(
s_{y}^{+}\gamma_{x}^{+}\!-\!s_{y}^{-}\gamma_{x}^{-} )  & - (  c_{x}%
^{+}\gamma_{y}^{+}\!+\!c_{x}^{-}\gamma_{y}^{-}\!+\!c_{y}^{+}\gamma_{x}^{+}\!+\!c_{y}%
^{-}\gamma_{x}^{-} )
\end{array}
\right]
\end{equation}
    \hrulefill
\end{figure*}
The following proposition describes the
asymptotic behavior of the Gramian channel matrix when $t_\mathrm{pol}=2$, where we must introduce some additional quantities here to formulate the result, namely\footnote{Each of these four quantities can be expressed as the product of the two sines of the two split angles from $\gamma_x^+$, $\gamma_x^-$, $\gamma_y^+$ and $\gamma_y^-$ that limit with each of the four edges of the tetrahedron in Fig.~\ref{fig:Angles}. }
\begin{align*}
\sigma^{++}   =\frac{\left(  L_{x}-x_{0}\right)  \left(  L_{y}-y_{0}\right)
}{\mathsf{d}_{++}^{2}}\,\,\, &  \sigma^{+-} =\frac{\left(  L_{x}-x_{0}\right)  \left(  L_{y}+y_{0}\right)
}{\mathsf{d}_{+-}^{2}}\\
\sigma^{-+}   =\frac{\left(  L_{x}+x_{0}\right)  \left(  L_{y}-y_{0}\right)
}{\mathsf{d}_{-+}^{2}}\,\,\,&  \sigma^{--}   =\frac{\left(  L_{x}+x_{0}\right)  \left(  L_{y}+y_{0}\right)
}{\mathsf{d}_{--}^{2}}.
\end{align*}

\begin{proposition} \label{prop:UPA2x3}
Assume $t_\mathrm{pol} =2$ and $ r_\mathrm{pol} = 3$. Under the same asymptotic conditions as in Proposition \ref{prop:UPA3x3}, the matrix $\mathbf{W}^{2\times3}$ converges to $\overline{\mathbf{W}}^{2\times3}$, the entries of which can be described as follows. Let $\rho = D^2/({32L_{x}L_{y}})$. The entries of this matrix take the form
\[
 \overline{\mathbf{W}}^{2\times3} \!=\!\!\! \left[
\begin{array}
[c]{ccc}
\frac{\Phi_{2}D^2}{2} & 0 & 0 \\
0 & \frac{\Phi_{2}D^2}{2} & 0 \\
0 & 0 & 0
\end{array} \right]
 + \rho\left[
\begin{array}
[c]{ccc}
 (\ast) & (\bullet)  &  (\bullet \bullet) \\
(\bullet) & (\ast \ast) &  (\bullet \ast)\\
(\bullet \bullet)  & (\bullet \ast) & (\square)
\end{array} \right]
\]
with
\begin{align*}
(\ast)= & -c_{y}^{+}\gamma_{x}^{+}+  \left(4-\left(  c_{x}^{+}\right)
^{2}\right)c_{x}^{+}\gamma_{y}^{+}+\left( 4-\left(c_{x}^{-}\right)
^{2}\right)c_{x}^{-}\gamma_{y}^{-}\\
&-c_{y}^{-}
\gamma_{x}^{-}+  \left(  s_{x}^{+}\right)  ^{2}\left(
\sigma^{++}+\sigma^{+-}\right)  +\left(  s_{x}^{-}\right)  ^{2}\left(
\sigma^{-+}+\sigma^{--}\right) \\
(\bullet)= &   4
\log\frac{\mathsf{d}_{--}\mathsf{d}_{++}}{\mathsf{d}_{+-}\mathsf{d}_{-+}
}- z_{0}^{2} \left(  \mathsf{d}_{++}^{-2}+\mathsf{d}
_{--}^{-2}-\mathsf{d}_{+-}^{-2}-\mathsf{d}_{-+}^{-2}\right)\\
(\bullet \bullet)=  &  \frac{\left(  s_{x}^{-}\right)  ^{3}}{{c_{x}^{-}}}\left(
{\sigma^{-+}+\sigma^{--}}\right) -  \frac{\left(
s_{x}^{+}\right)  ^{3}}{c_{x}^{+}}\left(  {\sigma^{++}+\sigma^{+-}}\right)  \\
&  +\left(s_{x}^{-}\right)^{3}\gamma_{y}^{-}-\left(s_{x}^{+}\right)^{3}\gamma_{y}^{+}\\
(\ast \ast)= & -c_{x}^{+}\gamma_{y}^{+}+  \left(4-\left(  c_{y}^{+}\right)
^{2}\right)c_{y}^{+}\gamma_{x}^{+}+\left( 4-\left(c_{y}^{-}\right)
^{2}\right)c_{y}^{-}\gamma_{x}^{-}\\
&-c_{x}^{-}
\gamma_{y}^{-}+  \left(  s_{y}^{+}\right)  ^{2}\left(
\sigma^{++}+\sigma^{-+}\right)  +\left(  s_{y}^{-}\right)  ^{2}\left(
\sigma^{+-}+\sigma^{--}\right) \\
(\bullet \ast)= & \frac{\left(  s_{y}^{-}\right)  ^{3}}{{c_{y}^{-}}}\left(
{\sigma^{+-}+\sigma^{--}}\right) -  \frac{\left(
s_{y}^{+}\right)  ^{3}}{c_{y}^{+}}\left(  {\sigma^{++}+\sigma^{-+}}\right)  \\
&  +\left(s_{y}^{-}\right)^{3}\gamma_{x}^{-}-\left(s_{y}^{+}\right)^{3}\gamma_{x}^{+}\\
(\square)= &    \left(  1+\left(  c_{y}^{+}\right)  ^{2}\right)
c_{y}^{+}\mathcal{\gamma}_{x}^{+}+\left(  1+\left(  c_{y}^{-}\right)
^{2}\right)  c_{y}^{-}\mathcal{\gamma}_{x}^{-} \\
& +  \left(  1+\left(  c_{x}^{+}\right)
^{2}\right)  c_{x}^{+}\mathcal{\gamma}_{y}^{(+)}+\left(  1+\left(  c_{x}
^{-}\right)  ^{2}\right)  c_{x}^{-}\mathcal{\gamma}_{y}^{(-)} \\
& - \left(  s_{x}^{+}\right)  ^{2}\left(
\sigma^{++}+\sigma^{+-}\right)  -\left(  s_{x}^{-}\right)  ^{2}\left(
\sigma^{-+}+\sigma^{--}\right)   \\
& - \left(  s_{y}^{+}\right)  ^{2}\left(
\sigma^{++}+\sigma^{-+}\right)  -\left(  s_{y}^{-}\right)  ^{2}\left(
\sigma^{+-}+\sigma^{--}\right)  
\end{align*}
where we recall that the trigonometric quantities $s_x^\pm,c_x^\pm,s_y^\pm,c_y^\pm$ are introduced in Proposition~\ref{prop:UPA3x3}. 

\end{proposition}
\begin{IEEEproof}
    See Appendix \ref{sec:UPA3x32x3}.
\end{IEEEproof}

Depending on the geometry of the scenario, one can slightly simplify the above equations. For example, if the UPA is placed symmetrically (so that $x_0=y_0 = 0$) 
the off-diagonal coefficients of the asymptotic Gramian matrix become zero, and $\overline{\mathbf{W}}^{3\times 3}$ in (\ref{eq:AsymGrammUPA3x3}) becomes diagonal.


\section{Numerical Analysis}
\label{sec:results}
This section provides an analysis of the benefits of XL-MIMO when multiple polarizations are employed. The first two subsections study the circumstances under which additional spatial degrees of freedom can be exploited by using the above asymptotic characterization. The optimum array dimensions are characterized for any given value of the reference SNR. Section \ref{sec:results_Nr}, on the other hand, analyzes the gains provided by increasing the number of receiving antennas and how this has an impact on the optimal transmitting aperture length. The receive antenna separation has an important role to attain the maximum spectral efficiency, and Section \ref{sec:results_Delta_RX} studies the connection between all these parameters. 

\subsection{Eigenvalue distribution}
\label{sec:results_eigenvalue}
\begin{figure}[htbp]
\centerline{\includegraphics[width=3.4 in, clip=true, trim={0 3 0 3}]{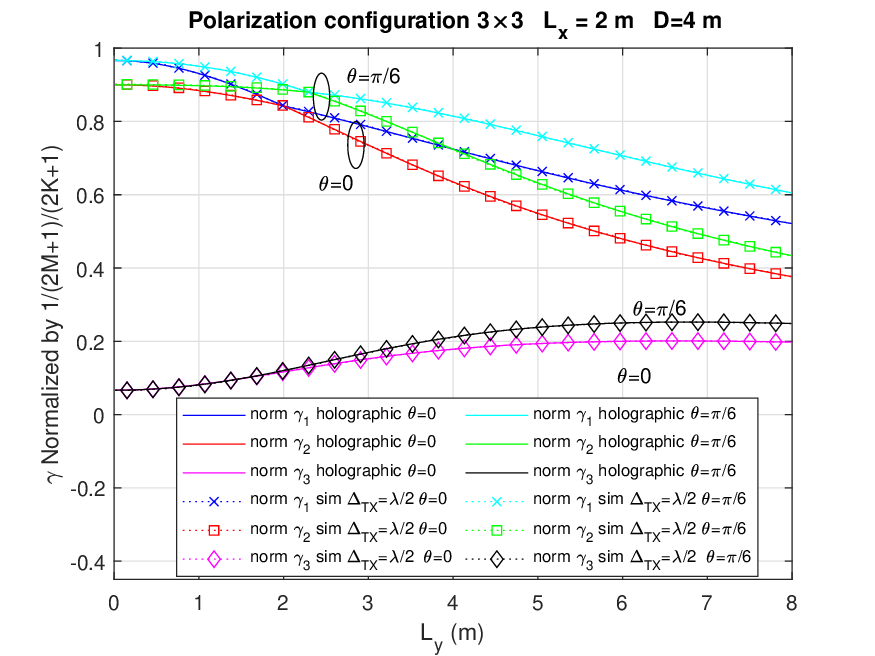}}
\caption{ Comparison of the expressions in the holographic regime with the simulated data. Magnitude of the eigenvalues of the Gramian in (\ref{eq:defW}) normalized by the total number of radiating elements $(2M+1)(2K+1)$ when a terminal is placed at $D=4$ m with elevation $\theta$.}
\label{fig:eigenvalues_3x3_UPA}
\end{figure}

The expressions obtained in Section \ref{sec:channel_rank} considered an infinitesimal antenna separation. In the following we will check the accuracy of the derived expressions and we will additionally analyze how the eigenvalues of the Gramian in (\ref{eq:defW}) depend on the antenna dimensions when $t_\mathrm{pol}=r_\mathrm{pol}=3$. 
Fig.~\ref{fig:eigenvalues_3x3_UPA} depicts the normalized magnitude of the eigenvalues with respect the number of radiating elements for an UPA of dimensions $2L_x\times 2L_y$, and a single element terminal at $D=4$~m with elevation $\theta \in \{0,\pi/6\}$, as a function of the dimension $L_y$, keeping $L_x=2$ fixed. 
The antenna separation was fixed to $\Delta_{T}=\lambda/2$ in this case, so that $L_y$ is directly proportional to the number of elements of the array.
We can observe that the analytical expressions obtained in the holographic regime (solid lines) are very accurate to the actual eigenvalues (dotted lines) for the different values of $\theta$ and $L_y$.  
Fig.~\ref{fig:eigenvalues_3x3_UPA} illustrates that the importance of the smallest channel eigenvalue increases when the aperture length is enlarged, although this comes at the cost of reducing the magnitude of the two largest eigenvalues. 

\subsection{Optimal antenna dimensions}
\begin{figure}[htbp]
\centerline{\includegraphics[width=3.4 in, clip=true, trim={0 3 0 3}]{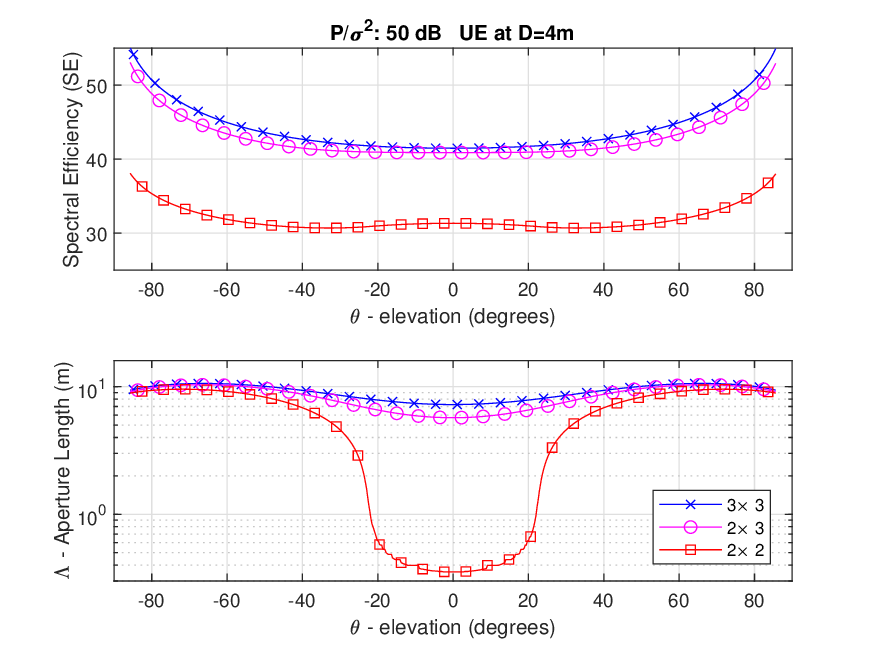}}
\caption{Top: Spectral Efficiency obtained for the different polarization configurations as function of the elevation $\theta$. Bottom: Optimal Aperture Length  ($\Lambda$) as a function of the elevation. Receiver at $D=4$ m.} 
\label{fig:SE_L_vs_elevation_ULA}
\end{figure}

In Section \ref{sec:ULA} we showed that, for a given value of the normalized SNR ($\mathsf{SNR}_0$), the achievable rate for the ULA case is dependent of two parameters, the ratio between antenna size and distance $\varrho=L/D$ and the elevation $\theta$. In this regard, Fig.~\ref{fig:SE_L_vs_elevation_ULA} represents the maximum achievable rate (top) and the optimum ULA length $\Lambda = 2L$ (bottom) as a function of the elevation ($\theta$) for a receiver placed at $D=4$ assuming three different polarization cases $(t_\mathrm{pol},r_\mathrm{pol}) \in \{(3,3),(2,3),(2,2)\}$. Notice that the optimum spectral efficiency tends to increase with the elevation because the receiver becomes closer to the ULA. 

\begin{figure}[htbp]
\centerline{\includegraphics[width=3.4 in, clip=true, trim={0 3 0 3}]{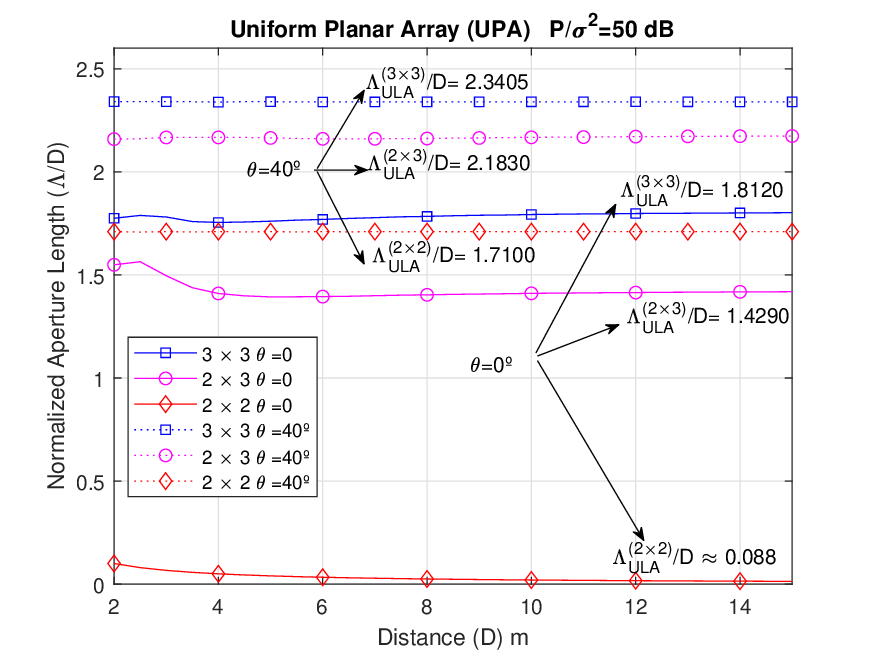}}
\caption{Normalized aperture length of UPA with respect the distance to the RX (D). 
receiver with elevation $\theta=0, \theta=40º$. Results compared with the aperture length derived when a ULA is considered.}
\label{fig:NormApertureLengthUPA_vs_D}
\end{figure}

Regarding the UPA configuration, it was also established that the achievable rate for a fixed reference $\mathsf{SNR}_0$ only depends on the relative geometry of the scenario. To further investigate this point, we consider the situation in which the user is centered with respect to the UPA ($x_0=y_0=0$). Given the above asymptotic analysis, one may conclude that the spectral efficiency can be seen as a function of the elevation angle $\theta$ and the ratio $\Lambda_{\mathrm{UPA}}/D$, where $\Lambda_{\mathrm{UPA}}=2\sqrt{L_x^2 + L_y^2}$ is the \textit{aperture length}, i.e. the largest separation between two antennas in given array \cite[Def. 4.1]{bjornson24book}.
Fig. \ref{fig:NormApertureLengthUPA_vs_D} represents the normalized aperture length $\Lambda/D$ that maximizes the spectral efficiency as a function of the distance to the receiver $D$ (which establishes the value of $\mathsf{SNR}_0$) for elevations $\theta \in \{0,40^\circ \}$ and different polarization configurations. It can be seen that the normalized aperture length shows an almost linear dependence with the distance $D$, with almost the same dependence for the UPA and ULA cases. 

Results depicted in Table \ref{tab:ratio_ULA_D} show that in order to attain the maximum spectral efficiency for polarization configuration $(3\times3)$ and $\theta=0$, the aperture length becomes $\Lambda \approx 1.8120\times D$, the aperture length is $\Lambda \approx 1.4290\times D$ for the configuration with $(2\times2)$. However, it is interesting to see that in case we would like to attain the 90\% of the maximum spectral efficiency (see Table \ref{tab:ratio_ULA_D_090}) the required aperture length is lower, reduced by a factor 3.7 at $D=4$ and 3.3 at $D=8$. 

\begin{table}[ht]

\begin{center}
\begin{tabular}{||c|c|c|c||} 
 \hline\hline
 \multicolumn{1}{|c|}{} &
 \multicolumn{3}{|c|}{max. SE}   \\
 \hline
 $\theta$ & $3\times3$ & $3\times2$ & $2\times2$  \\ 
 \hline\hline
  0 & 1.8120 & 1.4290 &  0.0885    \\ 
 \hline
 10 & 1.8510 & 1.4815 &  0.0995    \\
 \hline
 20 & 1.9660 & 1.6425 &  0.1545    \\
 \hline
 30 & 2.1405 & 1.8975 & 1.1475 \\
 \hline
 40 & 2.3405 & 2.1830 & 1.7100 \\ 
 \hline
 50 & 2.5180 & 2.4205 & 2.0730 \\ 
 \hline
 60 & 2.6305 & 2.5615 & 2.2960 \\ 
 \hline
 70 & 2.6430 & 2.5790 & 2.3875 \\ 
 \hline
 80 & 2.5170 & 2.4470 & 2.3390 \\ 
 
 \hline
\end{tabular}
\end{center}
\caption{Normalized aperture length $\Lambda/{D}$ for  different polarization configurations as a function of the elevation ($\theta$) in deg. to attain the maximum spectral efficiency (SE). $\frac{P}{\sigma^2} =50 \mathrm{dB} $. }\label{tab:ratio_ULA_D}
\end{table}

\begin{table}[ht]

\begin{center}
\begin{tabular}{||c|c|c|c|c|c|c||} 
 \hline\hline
 \multicolumn{1}{|c|}{} &
 \multicolumn{3}{|c|}{90\% max. SE ($D=4$)} &
 \multicolumn{3}{|c|}{90\% max. SE ($D=8$)}  \\
 \hline
 $\theta$ & $3\times3$ & $3\times2$ & $2\times2$  & $3\times3$ & $3\times2$ & $2\times2$ \\ 
 \hline\hline
  0 &  1.9400   & 1.5960  & 0.0020 &  4.3680   & 3.5880  & 0.0020  \\ 
 \hline
 10 &  2.0060   & 1.6720  & 0.0020 &  4.5140   & 3.7600  & 0.0020 \\
 \hline
 20 &  2.2220   & 1.9380  & 0.0020 &  4.9920 &   4.3520  &  0.0020\\
 \hline
 30 &  2.6480   & 2.5240  & 0.0020 &   5.9220   & 5.6300  & 0.0020 \\
 \hline
 40 &  3.3800   & 3.5920  & 0.0040 &   7.4600  & 7.8500   & 0.0060\\ 
 \hline
  
 \hline
\end{tabular}
\end{center}
\caption{ Smallest $\Lambda$ (m) to attain 90\% of the maximum SE for  different polarization configurations as a function of the elevation ($\theta$) in degrees. RX at $D\in \{4,8\}$ m. $\frac{P}{\sigma^2} =50 \mathrm{dB}$. }\label{tab:ratio_ULA_D_090}
\end{table}

\subsection{Increasing the number of receiving antennas}
\label{sec:results_Nr}
\begin{figure}[htbp]
\centerline{\includegraphics[width=3.5 in, clip=true, trim={0 3 0 3}]{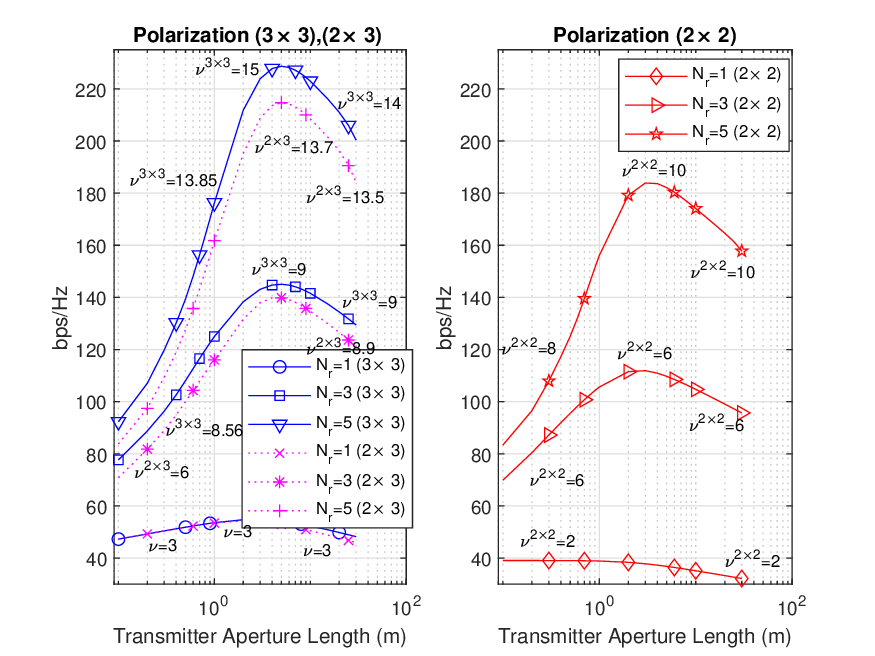}}
\caption{ Spectral Efficiency as a function of the transmitter aperture length for receive antenna elements $N_r={1,3,5}$ separated by $\Delta_{R}=\frac{\lambda}{2}$. Left: Polarization configuration $(3\times3),(2\times3)$. Right) Polarization configuration $(2\times2)$. The spatial DoF,  (\ref{eq:dof}), are included. Desired receiver at $D=3$ m. $M=20$. $\frac{P}{\sigma^2}=40 \mathrm{dB}$.}
\label{fig:SE_Nr_}
\end{figure}

When the number of receiving antennas is $N_r>1$ 
we must consider the generalized signal model by stacking the channel matrices in (\ref{eq:finalChannel}) for each of the receive antennas, see Remark \ref{rem:expand_signal_model}. The asymptotic expressions derived are no longer applicable, and we need to resort to numerical evaluation. Fig.~\ref{fig:SE_Nr_} represents the spectral efficiency in a scenario with a transmitter ULA and receiving antennas $N_r=1,3,5$ with the configurations $(3\times3), (2\times3)$ on the left and $(2\times2)$ on the right. 
In the same figure, we also indicate the attained spatial degrees of freedom (sDoF), $\nu^{t_\mathrm{pol}\times r_\mathrm{pol}}$, defined as the quotient in (\ref{eq:dof}), that provides information about the number of spatial streams that can be exploited in the communication. We observe that this number is dependent on the the aperture length, $\Lambda$, in certain cases it attains the total number of polarizations employed at the receiver side, $\nu=r_{\mathrm{pol}} \times N_r$. 
The plot on the left in Fig. \ref{fig:SE_Nr_} depicts the performance when the third polarization is exploited at the receiver side, i.e. the $(3\times3),(2\times3)$ configurations. The optimal dimensions were derived for $N_r=1$ in Table \ref{tab:ratio_ULA_D}, i.e. $\Lambda^{3\times3}=5.43$, $\Lambda^{2\times3}=4.28$ m, but for $N_r>1$ the aperture length for both configurations are almost the same, $\Lambda^{3\times3}=\Lambda^{2\times3}\approx5$~m, for $N_r \in \{3,5\}$. Furthermore, the degradation in terms of spectral efficiency for employing just 2 polarizations at the transmitter side (magenta lines) instead of 3 polarizations (blue lines) is small. Nevertheless, the attained number of DoF might differ from one configuration to another. For example, when $N_r \in \{1,3\}$ both configurations attain $\nu\in\{3,9\}$ DoF respectively with the optimal aperture length, but for $N_r=5$ the $\nu^{3 \times 3} \geq \nu^{2 \times 3}$, so the impact of the third transmitter polarization becomes relevant when increasing the number of receive antennas $N_r$. Let us remark that reducing the number of polarizations reduces the number of required RF chains, thus simplifying the transceiver.

\subsection{Receive antenna separation}
\label{sec:results_Delta_RX}
\begin{figure}[htbp]
\centerline{\includegraphics[width=3.4 in, clip=true, trim={0 3 0 3}]{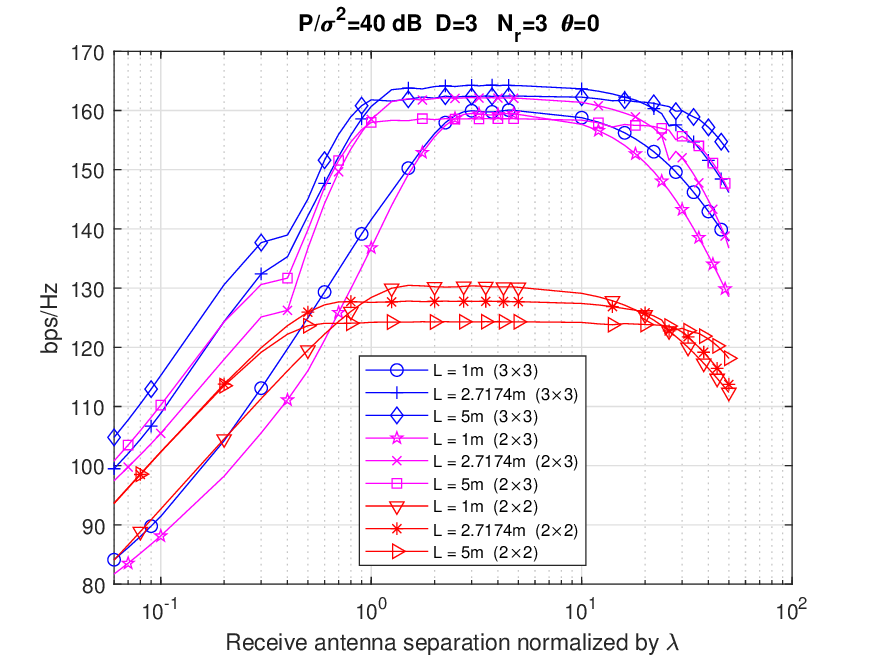}}
\caption{ Achievable rate as a function of the normalized separation of the receiving antennas ($\frac{\Delta_{R}}{\lambda}$), for different aperture lengths. $N_r=3$ receiving elements at $D=3$. $M=20$.}
\label{fig:modes_ULA_Rx_optim}
\end{figure}

We finally analyze here the sensitivity of the achievable rate as a function of the antenna antenna separation. To this end, 
Fig.~\ref{fig:modes_ULA_Rx_optim} depicts the achievable rate as a function normalized receive antenna separation, $\frac{\Delta_{R}}{\lambda}$, considering different polarizations for three  transmitter aperture lengths, $\Lambda=(1, 2.7174, 5)$ m, and $N_r=3$ when the receiver is placed at $D=3$~m. Notice that $\Lambda=2.7174$ was the optimal aperture length for the single element case $N_r=1$. The optimal spectral efficiency for each polarization configuration is attained for a given aperture length and receive antenna separation. In particular, for the ${2\times2}$ configuration the best performance is attained for $(\Lambda,\Delta_R)=(1, 1.2\lambda)$. On the other hand, for configurations ${3\times3}$ and ${3\times2}$ the  receive antennas should be separated by  $\Delta_{R}=1.5\lambda$ while the aperture length become $\Lambda^{3\times3}=\Lambda^{3\times2}=2.7174$ m. Notice that in case of reducing the aperture length $\Lambda^{3\times3}=\Lambda^{3\times2}=1$ keeping $\Delta_{R}=1.5\lambda$ the spectral efficiency degrades up to 10\%, but in case of adjusting the receive antenna separation to $\Delta_{R}=3\lambda$ the degradation is around 3\% in the particular case of Fig. \ref{fig:modes_ULA_Rx_optim}. Therefore, adjusting the receive antenna separation allows reducing the requirements on the aperture length at the cost of reducing the achievable rate.

\begin{figure}[htbp]
\centerline{\includegraphics[width=3.4 in, clip=true, trim={0 3 0 3}]{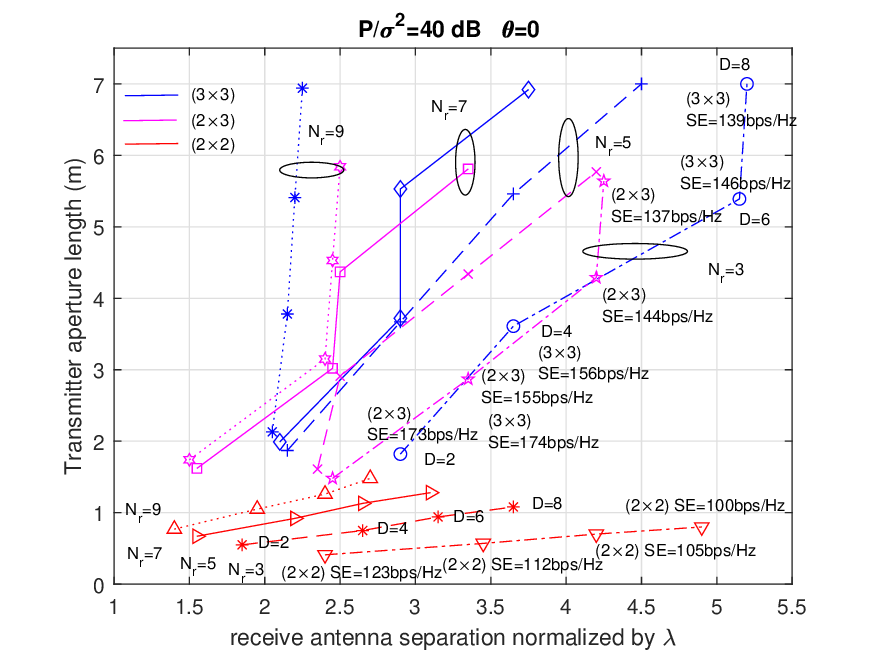}}
\caption{Transmitter aperture length ($\Lambda$) vs normalized receiver antenna separation by $\lambda$ for attaining the maximum spectral efficiency under each polarization configuration $(3\times3),(2\times3),(2\times2)$, receiver position $D={4,8}$ m, and $N_r={3,5,7,9}$. $\frac{P}{\sigma^2}=40$ dB, $\theta=0$, $M=20$.}
\label{fig:Dimensions_antennas_Nr_D_40dB}
\end{figure}

Fig.~\ref{fig:Dimensions_antennas_Nr_D_40dB} illustrates the dependence of aperture length ($\Lambda$) and receive antenna separation normalized by $\lambda$  when the receiver is placed at distances $D \in \{2, 4, 6, 8\}$ m, with elevation $\theta=0$ and equipped with $N_r \in \{3, 5, 7, 9\}$ antennas. For the case of $N_r=3$ we include the attained spectral efficiency at  different distances for each polarization configuration. The attained spectral efficiency for $N_r=\{5,7,9\}$ has not been included, but it has been observed that scales linearly with $N_r$. What can be concluded from Fig. \ref{fig:Dimensions_antennas_Nr_D_40dB} is that the optimal spectral efficiency is attained with a receive antenna separation that increases with the distance of the terminal ($D$). For example, with $N_r=3$ and $D=8$ m, the $\Delta_{R}^{(2\times2)}\approx 4.8\lambda$, while $\Delta_{R}^{(3\times3)}\approx 5.3\lambda$. 

\begin{figure}[htbp]
\centerline{\includegraphics[width=3.4 in, clip=true, trim={0 3 0 3}]{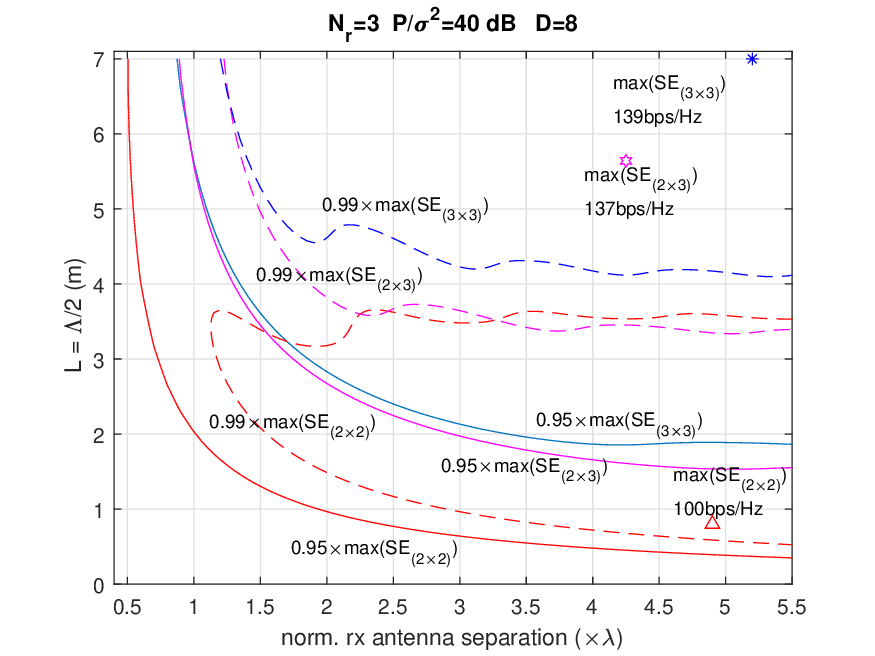}}
\caption{Transmitter aperture length vs normalized receiver antenna separation by $\lambda$ for attaining the 100\%, 99\% and 95\% of maximum spectral efficiency under the polarization configuration $(3\times3),(3\times2),(2\times2)$ and receiver at $D=8$ m. $N_r=3$. $\frac{P}{\sigma^2}=40$ dB. $\theta=0$.$M=20$.}
\label{fig:SE_region_dimensions}
\end{figure}

With the objective to elucidate how the requirements for the aperture length and receive antenna separation $(\Lambda,\Delta_R)$ relax if we are interested in attaining a percentage of maximum achievable rate, Fig. \ref{fig:SE_region_dimensions} illustrates the achievable rate as a function of $\Lambda$ and $\Delta_R$ when the receiver is placed at $D=8$ m and equipped with $N_r=3$ antennas. For configuration $(3\times3)$ the maximum achievable rate is obtained with $(\Lambda^{3\times 3},\Delta_R^{3 \times 3}) = (7,5.2\lambda)$, but there are multiple combinations to attain the 99\% of the maximum: $(\Lambda^{3\times 3},\Delta_R^{3 \times 3}) = (7,1.3\lambda)$,$(4.1,3\lambda)$, or the 95\% of the maximum: $(\Lambda^{3\times 3},\Delta_R^{3 \times 3}) = (3,2\lambda)$,$(2,3.5\lambda)$. Similar conclusions can be derived for the other polarization configurations. In general, the gains in terms of spectral efficiency provided by the XL-MIMO are not too sensitive to variations in the antenna dimensions.


\section{Conclusion}
\label{sec:conclusion}
Additional spatial degrees of freedom are possible in XL-MIMO configurations thanks to the use of antennas with three polarizations either at the transmitter and/or at the receiver side. These gains are available when the wireless communication takes places in the {near-field}, otherwise the conventional dual-polarized antennas are already optimal. We have provided a framework that allows to describe the Gramian of the channel in the {holographic} regime, showing that eigenvalues of the channel, and the spectral efficiency, are dependent of the ratio of the aperture length of the UPA and the distance, $\frac{\Lambda}{D}$, and the elevation $\theta$. Likewise, the requirements for the transmitter aperture length can be relaxed by exploiting additional receive antennas and its separation. 
Finally, the most important place to deploy antenna elements with three polarizations is at the receiver side. Having them at the transmitter side becomes relevant $N_r \geq 3 $. This an important aspect, because we can reduce the complexity in terms of RF chains at the UPA. Furthermore, the optimal dimension of the UPA can be reduced if the separation of the antennas at the received side can be adjusted.


%

\appendices

\section{Proof of Lemma \ref{lemma:radiative}}
\label{sec:Ap_radiative}
It is sufficient to prove the result with $r_{\mathrm{pol}}=3$, since for
lower values of the number of receive polarization the matrix inside the
spectral norm is a submatrix of the one obtained in the case $r_{\mathrm{pol}
}=3$ (and has therefore lower spectral norm).\ Observe that we can express
$\mathcal{H}_{k,m}$ shown in (\ref{eq:channelModel}) as
\[
\mathcal{H}_{k,m}=\mathbf{P}_{k,m}^{\perp}+\left(  \frac{\mathrm{j}\lambda
}{2\pi r_{k,m}}-\frac{\lambda^{2}}{(2\pi r_{k,m})^{2}}\right)  \mathbf{\tilde
{P}}_{k,m}
\]
where $
\mathbf{\tilde{P}}_{k,m}=\mathbf{I}_{3}-3{\mathbf{r}_{k,m}\mathbf{r}
_{k,m}^{H}}/{\left\Vert \mathbf{r}_{k,m}\right\Vert ^{2}}$.
Let $\mathcal{S}_{t_{\mathrm{pol}}}$ denote a $3\times3$ selection matrix with
ones in the first $t_{\mathrm{pol}}$ diagonal elements and zeros elsewhere. We
can write 
\begin{align*}
\mathbf{E}_{k,m}^{t_{\mathrm{pol}}\times r_{\mathrm{pol}}} &  
=\left\vert \frac{\xi}{\lambda}\right\vert ^{2}\frac{\mathrm{j}\lambda
}{2\pi r_{k,m}^{3}}\left(  1+\frac{\mathrm{j}\lambda}{2\pi r_{k,m}}\right)
\mathbf{P}_{k,m}^{\perp}\mathcal{S}_{t_{\mathrm{pol}}}\mathbf{\tilde{P}}
_{k,m}\\
&  +\left\vert \frac{\xi}{\lambda}\right\vert ^{2}\frac{-\mathrm{j}\lambda
}{2\pi r_{k,m}^{3}}\left(  1-\frac{\mathrm{j}\lambda}{2\pi r_{k,m}}\right)
\mathbf{\tilde{P}}_{k,m}\mathcal{S}_{t_{\mathrm{pol}}}\mathbf{P}_{k,m}^{\perp
}\\
&  +\left\vert \frac{\xi}{\lambda}\right\vert ^{2}\frac{\lambda^{2}}{\left(
2\pi\right)  ^{3}r_{k,m}^{4}}\left\vert 1+\frac{\mathrm{j}\lambda}{2\pi
r_{k,m}}\right\vert ^{2}\mathbf{\tilde{P}}_{k,m}\mathcal{S}_{t_{\mathrm{pol}}
}\mathbf{\tilde{P}}_{k,m}.
\end{align*}
Now, using the fact that $\left\Vert \mathbf{AB}\right\Vert \leq\left\Vert
\mathbf{A}\right\Vert \left\Vert \mathbf{B}\right\Vert $, $\left\Vert
\mathbf{A+B}\right\Vert \leq\left\Vert \mathbf{A}\right\Vert +\left\Vert
\mathbf{B}\right\Vert $, $\left\Vert \mathbf{P}_{k,m}^{\perp}\right\Vert
\leq1$, $\left\Vert \mathcal{S}_{t_{\mathrm{pol}}}\right\Vert =1$,$\left\Vert
\mathbf{\tilde{P}}_{k,m}\right\Vert \leq4$ and Jensen's inequality we find
\begin{align*}
\left\Vert \mathbf{E}_{k,m}^{t_{\mathrm{pol}}\times r_{\mathrm{pol}}
}\right\Vert  &  \leq8\left\vert \frac{\xi}{\lambda}\right\vert ^{2}
\frac{\lambda}{2\pi r_{k,m}^{3}}\left(  1+\frac{\lambda}{2\pi r_{k,m}}\right)
\\
&  +32\left\vert \frac{\xi}{\lambda}\right\vert ^{2}\frac{\lambda^{2}}{\left(
2\pi\right)  ^{3}r_{k,m}^{4}}\left(  1+\left(  \frac{\lambda}{2\pi r_{k,m}
}\right)  ^{2}\right)  .
\end{align*}
Finally, observing that $r_{k,m}^{2}\geq d_{\inf}^{2}$ we obtain the first result
of Lemma \ref{lemma:radiative}. When $t_{\mathrm{pol}}\!=\!3$ we have $\mathcal{S}_{t_{\mathrm{pol}
}}\!=\!\mathbf{I}_{3}$ and thus  $\mathbf{P}_{k,m}^{\perp}\mathcal{S}
_{t_{\mathrm{pol}}}\mathbf{\tilde{P}}_{k,m}=\mathbf{P}_{k,m}^{\perp
}\mathcal{\ }$, while $\mathbf{E}_{k,m}^{t_{\mathrm{pol}}\times r_{\mathrm{pol}}}$
takes the form
\begin{align*}
\mathbf{E}_{k,m}^{t_{\mathrm{pol}}\times r_{\mathrm{pol}}} &  =-\left\vert
\frac{\xi}{\lambda}\right\vert ^{2}\frac{\lambda^{2}}{2\pi^{2}r_{k,m}^{4}
}\mathbf{P}_{k,m}^{\perp}\\
&  +\left\vert \frac{\xi}{\lambda}\right\vert ^{2}\frac{\lambda^{2}}{\left(
2\pi\right)  ^{3}r_{k,m}^{4}}\left[  1+\left(  \frac{\lambda}{2\pi r_{k,m}
}\right)  ^{2}\right]  \mathbf{\tilde{P}}_{k,m}\mathbf{\tilde{P}}_{k,m}.
\end{align*}
The lemma is proved by using the same bounds as above.

\section{Proof of Propositions \ref{prop:ULA3x3} and  \ref{prop:ULA2x3}} \label{sec:proofsPropsULA}
A direct evaluation of (\ref{eq:norm_W}) shows that the matrices $\mathbf{W}^{3\times 3}$ and $\mathbf{W}^{2\times 3}$ take the same expression as the matrices $\overline{\mathbf{W}}^{3\times 3}$ and $\overline{\mathbf{W}}^{2\times 3}$ in the statement of the two propositions, but simply replacing $\psi_k$, $k=2,\ldots,6$ by the quantities $s_M^{(k)}$ defined as
\begin{equation} \label{eq:defSk}
s_{M}^{(k)}  = \left\{ 
\begin{array}
[c]{cc}
\frac{1}{2M+1}\sum_{m=-M}^{M}{r_{m}^{-k}} & \text{
(k even)} \\
\frac{1}{2M+1}\sum_{m=-M}^{M}{ \frac{\left(  m\Delta_{T}
-D_s\right)}{ r_{m}^{k+1}}  } & \text{ (k odd)}.
\end{array}
\right.
\end{equation} 
with $r_m^2 = r_{0,m}^2= (m\Delta_T-y_0)^2 + z_0^2$. 
Hence, to prove these two propositions we only need to show that $\lim_{ M\Delta_T\to L}{s_{M}^{(k)}}=\psi_{k}$. The case $k=2$ directly follows from the definition of Riemann integral, that is
\begin{multline*}
    s_M^{(2)}  = \frac{1}{2M+1}\sum_{m=-M}^{M}\frac{  1}{ \left(  m\Delta_{T}-D_s\right)
^{2}+ D_c^{2} } \\ 
   = \frac{1}{\left(  2M+1\right)  \Delta_{T}}\int_{-M\Delta_{T}}^{M\Delta_{T}}\frac{1}{\left(  x-D_s\right)  ^{2}+D_c^{2}}dx+o(1) \\
     = \frac{1}{2LD_c}\left[  \arctan\left(  \frac{L+D_s}
{D_c}\right)  +\arctan\left(  \frac{L-D_s}{D_c}\right)  \right]  +o(1).
\end{multline*}
The convergence of $s_M^{(3)}$ and $s_M^{(4)}$ follows by the same type of Riemann integral approximation argument, together with the primitives \cite[2.148]{Gradshteyn}
\begin{align*}
    \int\!\!\frac{2ydy}{\left(  1\!+\!y^{2}\right)  ^{2}}  =\frac{-1}{1+y^{2}}, 
    \int\!\!\frac{dy}{\left(  1\!+\!y^{2}\right)  ^{2}}  =\frac{1}{2}\frac{y}{\left(1\!+\!y^{2}\right)  }\!+\!\frac{\arctan y}{2}.
\end{align*}
Regarding the convergence of $s_M^{(5)}$, the limit follows again from the definition of Riemann integral, together with 
\[
\int\frac{y}{\left(  1+y^{2}\right)  ^{3}}dy=-\frac{1}{4}\frac{1}{\left(
1+y^{2}\right)  ^{2}}.
\]
As for $s_{M}^{(6)}$, we operate in the same way using \cite[2.148]{Gradshteyn}
\[
\int\frac{1}{\left(  1+y^{2}\right)  ^{3}}dy=\frac{1}{4}\frac{y}{\left(
1+y^{2}\right)  ^{2}}+\frac{3}{8}\left[  \frac{y}{1+y^{2}}+\arctan y\right].
\]

\section{Proof of Propositions \ref{prop:UPA3x3} and \ref{prop:UPA2x3}} \label{sec:UPA3x32x3}

A direct evaluation shows that we can write
\begin{equation} \label{eq:ProjectionUPA}
\mathbf{P}_{k,m}^{\perp}=\frac{1}{r_{k,m}^{2}}\left[
\begin{array}
[c]{ccc}
\Delta_m^2+z_{0}^{2} & -\Delta_k \Delta_m & z_{0}\Delta_k\\
-\Delta_k \Delta_m  & \Delta_k^{2}+z_{0}^{2} & z_{0}\Delta_m \\
z_{0}\Delta_k & z_{0}\Delta_m & \Delta_k^{2}+\Delta_m^{2}
\end{array}
\right]
\end{equation}
where, with some abuse of notation, we have used the short hand notation $\Delta_k = k\Delta_T - x_0$ and $\Delta_m = m\Delta_T - y_0$. 
We can establish the convergence of the different terms in ${\mathbf{W}}^{3 \times 3}$ and ${\mathbf{W}}^{2 \times 3}$ by using here again the fact that for two continous functions $f(y)$, $g(x)$ and $n \in \mathbb{N}$, we have 
\begin{multline*}
\frac{1}{2M+1}\frac{1}{2K+1}\sum_{m=-M}^{M}\sum_{k=-K}^{K}\frac{f(k\Delta
_{T})g(m\Delta_{T})}{r_{k,m}^{2n}} \rightarrow \\ \rightarrow \frac{1}{2L_{x}}\frac{1}{2L_{y}}
\int_{-L_{y}}^{L_{y}}\int_{-L_{x}}^{L_{x}}\frac{f(x)g(y)}{R^{2n}(x,y) 
}dxdy.
\end{multline*}
where 
$
R^2(x,y) = \left(
x-x_{0}\right)  ^{2}+\left(  y-y_{0}\right)  ^{2}+z_{0}^{2}
$.
In particular, it directly follows from the above equation that 
\begin{multline*}
\frac{1}{N_{M,K}}\sum_{k,m} r_{k,m}^{-2} 
\rightarrow  \frac{1}{4 L_x L_y}
\!\!\int_{-L_{y}}^{L_{y}} \!\!\int_{-L_{x}}^{L_{x}} \!\!R^{-2}(x,y) dx dy =\Phi_2
\end{multline*}
with $N_{M,K}=(2M+1)(2K+1)$.
We can use the same approach on the rest of the terms in ${\mathbf{W}}^{3 \times 3}$, which by a direct Riemann approximation trivially converge to the corresponding double integrals as shown in (\ref{eq:AsymGrammUPA3x3DoubleIntegral}) at the top of the next page. At this point, it remains to evaluate the double integral of the different terms. 

The integration with respect one of the variables (say $y$) can be carried out by simply using fraction integration. In order to present the result, we need to introduce some additional functions as follows. With some abuse of notation, consider the function of one-variable $R(x)$ which is obtained by forcing $y=0$ in the equation above, that is $R(x)=R(x,0)$.
Likewise, we define the function $\psi_2(x)$ as a generalization of the quantity in (\ref{eq:DefPsi2}) for ULAs, namely
\begin{multline*}
    \psi_{2}(x) =
\frac{1}{2L_{y}
\sqrt{\left(  x-x_{0}\right)  ^{2}+z_{0}^{2}}} \times \\\Bigg[  \arctan
\frac{L_{y}-y_{0}}{\sqrt{\left(  x-x_{0}\right)  ^{2}+z_{0}^{2}}}
+\arctan  \frac{L_{y}+y_{0}}{\sqrt{\left(  x-x_{0}\right)  ^{2}
+z_{0}^{2}}}  \Bigg] 
\end{multline*}
so that we clearly have $\psi_{2}=\psi_{2}(0)$. Furthermore, one can easily see that $\Phi_2 = (2L_x)^{-1}\int_{-L_x}^{L_x} \psi_2(x)dx$ so that one can numerically evaluate $\Phi_2$ using a single integral. 
Finally,
$\psi_3(x)$ and $\psi_4(x)$ are defined as 
\begin{align*}
\psi_{3}(x)  &  =\frac{-y_{0}}{\left(  L_{y}^{2}+R^{2}(x)\right)  ^{2}-\left(  2L_{y}
y_{0}\right)  ^{2}} \\
\psi_{4}(x) &  =\frac{1/2}{ R^2(x)- y_0^2 } \left[
\frac{L_{y}^{2} +R^2(x) -2y_{0}^{2}}{\left(L_{y}^{2}+R^{2}(x)\right)  ^{2}-\left(  2L_{y}y_{0}\right)  ^{2}} + \psi_{2}(x) \right]
\end{align*}
so that here again we have $\psi_3=\psi_3(0)$ and $\psi_4=\psi_4(0)$. Using these definitions, we can express the matrix after carrying out the integral with respect to the $y$ variable as in (\ref{eq:AsymGrammUPA3x3SingleIntegral}) at the top of the next page. 

It therefore remains to compute the integrals with respect to the $x$ variable in the different entries of the matrix $\overline{\mathbf{W}}^{3\times3}$. The integrals of the $(1,2)$th and $(2,3)$th terms only involve the function $\psi_3(x)$ and can be integrated with little effort. The integral of the $(1,3)$th term involves $\psi_4(x)$ which is a bit more difficult to integrate due to the presence of the arctangent in  $\psi_2(x)$. However, one can easily find the primitive (for  $a \neq 0$)
\begin{multline} \label{eq:intarc32t}
   \!\!\! \int\frac{t}{\left(  1+t^{2}\right)  ^{3/2}}\arctan\left(  \frac{a}
{\sqrt{1+t^{2}}}\right)  dt=   \\ 
\!\!\! \!=\frac{1}{2a}\log\left(  1+\frac{a^{2}}{1\!+\!t^{2}}\right) \! - \! \frac{1}{\sqrt{1\!+\!t^{2}}}\arctan\left(  \frac{a}{\sqrt{1\!+\!t^{2}}
}\right) 
\end{multline}
(up to a constant) from which the integral of this term directly follows. Regarding the diagonal terms, they can be computed in a similar manner using again the primitive in (\ref{eq:intarc32t}) when integrating quantities that depend on the arctangent through $\psi_2(x)$ (details are omitted). The expression in (\ref{eq:AsymGrammUPA3x3}) is therefore obtained by simply using the trivial trigonometric identities (cf. Fig.~\ref{fig:Angles})
\begin{align*}
\frac{L_{x}-x_{0}}{\sqrt{\left(  L_{x}-x_{0}\right)  ^{2}+z_{0}^{2}}}  &
=\cos\mathcal{\beta}_{x}^{+},\,\frac{L_{x}+x_{0}}{\sqrt{\left(
L_{x}+x_{0}\right)  ^{2}+z_{0}^{2}}}=\cos\mathcal{\beta}_{x}^{-}\\
\frac{L_{y}-y_{0}}{\sqrt{\left(  L_{y}-y_{0}\right)  ^{2}+z_{0}^{2}}}  &
=\cos\mathcal{\beta}_{y}^{+},\,\frac{L_{y}+y_{0}}{\sqrt{\left(
L_{y}+y_{0}\right)  ^{2}+z_{0}^{2}}}=\cos\mathcal{\beta}_{y}^{-}
\end{align*}
together with their equivalents for the sines. 
\begin{figure*}[t]
    \centering
    \normalsize
\begin{equation}  \label{eq:AsymGrammUPA3x3DoubleIntegral}
\!\!\overline{\mathbf{W}}^{3\times3}
\!=\!\frac{D^2}{4 L_{x}L_{y}}\int_{-L_{y}}^{L_{y}}\int_{-L_{x}}^{L_{x}
}\frac{1}{R^{4}(x,y)}\left[
\begin{array}
[c]{ccc}
\left(  y-y_{0}\right)  ^{2}+z_{0}^{2} & -\left(  x-x_{0}\right)  \left(
y-y_{0}\right)   & z_{0}\left(  x-x_{0}\right)  \\
-\left(  x-x_{0}\right)  \left(  y-y_{0}\right)   & \left(  x-x_{0}\right)
^{2}+z_{0}^{2} & z_{0}\left(  y-y_{0}\right)  \\
z_{0}\left(  x-x_{0}\right)   & z_{0}\left(  y-y_{0}\right)   & \left(
x-x_{0}\right)  ^{2}+\left(  y-y_{0}\right)  ^{2}
\end{array}
\right]  dxdy
\end{equation}
\begin{equation}
\label{eq:AsymGrammUPA3x3SingleIntegral}
    \overline{\mathbf{W}}^{3\times3}=
    \frac{D^2}{2 L_{x}}\int_{-L_{x}}^{L_{x}}\left[
\begin{array}
[c]{ccc}%
\psi_{2}(x)-\left(  x-x_{0}\right)  ^{2}\psi_{4}(x) & -\left(  x-x_{0}\right)
\psi_{3}(x) & z_{0}\left(  x-x_{0}\right)  \psi_{4}(x)\\
-\left(  x-x_{0}\right)  \psi_{3}(x) & \left(  \left(  x-x_{0}\right)
^{2}+z_{0}^{2}\right)  \psi_{4}(x) & z_{0}\psi_{3}(x)\\
z_{0}\left(  x-x_{0}\right)  \psi_{4}(x) & z_{0}\psi_{3}(x) & \psi
_{2}(x)-z_{0}^{2}\psi_{4}(x)
\end{array}
\right]  dx
\end{equation}
    \begin{equation}
        \label{eq:AsymGrammUPA2x3}
\!\!\!\overline{\mathbf{W}}^{2 \times 3} \!\! = \!\frac{D^2}{2L_{x}}\!\!
\int_{-L_{x}}^{L_{x}}\!\!\left[
\begin{array}
[c]{ccc}
\psi_{2}(x)-\Delta_{x}^2 \left(  \psi_{4}(x)+z_{0}^{2}\psi
_{6}(x)\right)  & -\Delta_{x}  \left(  \psi_{3}(x)+z_{0}^{2}
\psi_{5}(x)\right)  & \Delta_{x}  z_{0}^{3}\psi_{6}(x)\\
-\Delta_{x}  \left(  \psi_{3}(x)+z_{0}^{2}\psi_{5}(x)\right)  &
\Delta_{x}^{2} \left(\psi_{4}(x)+ z_{0}^{2}\psi_{6}(x) \right) + z_0^4 \psi_{6}(x)
& z_{0}^{3}\psi_{5}(x)\\
\Delta_{x} z_{0}^{3}\psi_{6}(x) & z_{0}^{3}\psi_{5}(x) &
z_{0}^{2}\left[  \psi_{4}(x)-z_{0}^{2}\psi_{6}(x) \right]
\end{array}
\!\! \right] \!  dx
    \end{equation}    



    \hrulefill
    \end{figure*}
Let us now focus on the convergence of the Gramian matrix in (\ref{eq:norm_W}) when $t_\mathrm{pol} = 2$, we simply observe that we can write the matrix $\mathbf{W}^{2 \times 3}$  as 
\[
{\mathbf{W}}^{2 \times 3} = \frac{D^2}{(2M+1)(2K+1)} \sum_{k=-K}^{K} \sum_{k=-M}^{M} \frac{1}{r^6_{k,m}} \mathcal{T}^{2 \times 3}_{k,m} 
\]
where $\mathcal{T}^{2 \times 3}_{k,m}$ is the following matrix 
\[
 \mathcal{T}^{2 \times 3}_{k,m} = \\
 = \left[
\begin{array}
[c]{ccc}
 (\ast) & (\bullet)  & z_0 \Delta_k \\
(\bullet) & (\ast \ast) & z_{0}^{3}\Delta_m\\
z_{0}^{3} \Delta_k  & z_{0}^{3}\Delta_m & z_{0}^{2}\left( r_{k,m}^2-z_{0}^{2}\right)
\end{array} \right]
\]
with the missing entries defined as
\begin{align*}
 (\ast) & = \left(\Delta_m^2 + z_0^2\right)^2 +\Delta_k^2 \Delta_m^2 = r_{k,m}^4 - \Delta_k^2 \left(r_{k,m}^2 + z_0^2 \right)  \\
  (\ast \ast) & = \left(\Delta_k^2 + z_0^2\right)^2 +\Delta_k^2 \Delta_m^2 
  = \Delta_k^2 \left( r_{k,m}^2 + z_0^2 \right) + z_0^4 \\
    (\bullet) & =  -\Delta_k\Delta_m(r_{k,m}^2+z_0^2)
\end{align*}
Having expressed the entries of this matrix in this form, it only remains to use the Riemann integral definition in each of these terms to see that they individually converge to the corresponding double integral. Operating in the same way as in the previous case, we can then carry out the integration with respect to the $y$ variable, the only difficulty coming from terms with denominator equal to $R^6(x,y)$. These can be solved by noticing that 
\[
\int\frac{1}{\left(  1+t^{2}\right)  ^{3}}dt=\frac{1}{4}\frac{t}{(1+t^{2}%
)^{2}}+\frac{3}{8}\frac{t}{(1+t^{2})}+\frac{3}{8}\arctan t
\]
up to a constant. Using the above primitive we can easily carry out the integration with respect to $y$ and directly obtain the expression of $\overline{\mathbf{W}}^{2 \times 3}$ in (\ref{eq:AsymGrammUPA2x3}) at the top of the page. In this expression, we have additionally introduced the two functions
\begin{align*}
& \psi_{5}(x)  =\frac{-y_{0}\left(  R^{2}(x)+L_{y}^{2}\right)  }{\left[  \left(
R^{2}(x)+L_{y}^{2}\right)  ^{2}-\left(  2L_{y}y_{0}\right)  ^{2}\right]  ^{2}} \\
& \psi_{6}(x)   =\frac{1/4}{R^2(x)-y_0^2}
\frac{\left(  R^{2}(x)+L_{y}^{2}\right)  ^{2}
-4y_{0}^{2}R^{2}(x)}{\left(  \left(  R^{2}(x)+L_{y}^{2}\right)  ^{2}-\left(2L_{y}y_{0}\right)  ^{2}\right)  ^{2}} \\ 
& +\frac{3/8}{(R^2 (x)-y_0^2)^2}\left[  \frac{R^{2}(x)-y_{0}^{2}}{\left(  R^{2}(x)+L_{y}^{2}\right)  ^{2}-\left(  2L_{y}y_{0}\right)  ^{2}}+\psi_{2}(x)\right] 
\end{align*}
and we introduced the short-hand notation $\Delta_x = x-x_0$. 

To finish the proof of Proposition~\ref{prop:UPA2x3}, it remains to compute the different integrals with respect to the remaining variable $x$. The integral of the $(1,2)$th term can easily be computed by using partial fraction decomposition. The same procedure can be used to find a closed form expression for the $(2,3)$th term, which additionally needs the primitive
\begin{equation}
\int\frac{1}{\left(  1+t^{2}\right)  ^{2}}dt=\frac{1}{2}\left[  \frac
{t}{1+t^{2}}+\arctan t\right]  \label{eq:integralsquareden}.
\end{equation}
The rest of the terms depend on the arctangent through $\psi_2(x)$,  $\psi_4(x)$ and $\psi_6(x)$ and require some more work. In particular, the integrals involving these expressions can be solved by using (\ref{eq:intarc32t}) and the additional primitive (for $a\neq 0$)
\begin{multline}
\int\frac{1}{\left(  1+t^{2}\right)  ^{5/2}}\arctan\left(  \frac{a}
{\sqrt{1+t^{2}}}\right)  dt =\\=\frac{2a^{2}-1}{3a^{3}}\sqrt{1\!+\!a^{2}}
\arctan\left(  \frac{t}{\sqrt{1\!+\!a^{2}}}\right) -\frac{t}{6a\left(
1\!+\!t^{2}\right)  } +\\ +\frac{\left(  2t^{2}
\!+\!3\right)  t}{3\left(  1\!+\!t^{2}\right)  ^{3/2}}\arctan\left(  \frac{a}
{\sqrt{1\!+\!t^{2}}}\right) -\frac{3a^{2}\!-\!2}{6a^{3}}\arctan t.
\end{multline}
The above formulas can be used in combination with partial fraction decomposition to establish a closed form expression for the integrals in (\ref{eq:AsymGrammUPA2x3}), which can be shown to coincide with those in the statement of Proposition~\ref{prop:UPA2x3}. 



\ifCLASSOPTIONcaptionsoff
  \newpage
\fi



%

\bibliographystyle{IEEEtran}
\bibliography{./biblio}








\end{document}